\documentclass[english,format=acmsmall, review=false, screen=true]{article}
\usepackage[T1]{fontenc}
\usepackage[latin9]{inputenc}
\usepackage{verbatim}
\usepackage{slashed}
\usepackage{amsmath}
\usepackage{amssymb}

\makeatletter
\@ifundefined{date}{}{\date{}}
\usepackage{graphicx}
\usepackage{subfig}
\usepackage{multicol}

\makeatother

\usepackage{babel}
\begin{document}

\title{Efficient Detection of Complex Event Patterns Using Lazy Chain Automata}
\maketitle \begin{multicols}{2}
\begin{center} Ilya Kolchinsky\\ Technion, Israel Institute of Technology \par\end{center}
\begin{center} Assaf Schuster\\ Technion, Israel Institute of Technology \par\end{center}
\end{multicols}
\begin{abstract}
Complex Event Processing (CEP) is an emerging field with important
applications in many areas. CEP systems collect events arriving from
input data streams and use them to infer more complex events according
to predefined patterns. The Non-deterministic Finite Automaton (NFA)
is one of the most popular mechanisms on which such systems are based.
During the event detection process, NFAs incrementally extend previously
observed partial matches until a full match for the query is found.
As a result, each arriving event needs to be processed to determine
whether a new partial match is to be initiated or an existing one
extended. This method may be highly inefficient when many of the events
do not result in output matches.

We propose a lazy evaluation mechanism that defers processing of frequent
event types and stores them internally upon arrival. Events are then
matched in ascending order of frequency, thus minimizing potentially
redundant computations. We introduce a lazy Chain NFA, which utilizes
the above principle, and does not depend on the underlying pattern
structure. An algorithm for constructing a Chain NFA for common pattern
types is presented, including conjunction, negation and iteration.
In addition, we propose a Tree NFA that does not require the frequencies
of the event types to be defined in advance. Finally, we experimentally
evaluate our mechanism on real-world stock trading data. The results
demonstrate a performance gain of two orders of magnitude over traditional
NFA-based approaches, with significantly reduced memory resource requirements.
\end{abstract}

\section{Introduction}

\label{sec:Introduction}

The goal of complex event processing systems is to efficiently detect
complex patterns over streams of events. A CEP engine is responsible
for filtering and combining primitive events into higher-level, \textit{complex
events}, which are then reported to end users. Areas in which CEP
is widely applied include financial services \cite{DemersGHRW06},
RFID-based inventory management \cite{WangL05}, and electronic health
record systems \cite{HaradaH05}.

The patterns recognized by CEP systems are normally created using
declarative specification languages. They are usually based on relational
languages extended with additional operators, making it possible to
define a wide range of pattern types, such as sequences, disjunctions
and iterations. Basic operators can be combined into arbitrarily composite
expressions. Filters, constraints, time windows, and mutual conditions
between events can be applied. Examples of complex event specification
languages include SASE \cite{WuDR06}, CQL \cite{ArasuBW06}, CEL
\cite{BrennaDGHOPRTW07} and CEDR \cite{BargaGAH07}.

Different patterns require different policies as to whether a primitive
event selected for some pattern match can be considered again for
future matches, a notion known as \textit{event selection strategy}.
In this paper we assume that, for any pattern, all possible matching
combinations of events are requested to be detected. Consequently,
a primitive event is allowed to participate in an unlimited number
of matches. This selection strategy is called \textit{skip-till-any-match}\cite{AgrawalDGI2008}.

To illustrate the above notions, consider the following example.

\newtheorem{example}{Example}
\begin{example}

A stock market monitoring application is requested to detect non-standard
behavior of stock prices. For each stock identifier, a primitive event
is generated when its price exceeds some predefined value. We would
like to detect a complex event in which an irregularly high price
of stock \textit{St1} is detected, followed by a high price of stock
\textit{St2}, which is also followed by a high price of stock \textit{St3},
within a time window of one hour. High price is defined as exceeding
a given threshold \textit{T.}

\end{example}

Now, consider an input stream $a_{1},a_{2},b_{1},b_{2},c$, where
events \textit{a}, \textit{b} and \textit{c} denote observations of
high values of stocks \textit{St1}, \textit{St2} and \textit{St3}
respectively. Under the skip-till-any-match selection strategy, the
matches to be returned are $\left\{ a_{1}b_{1}c\right\} $, $\left\{ a_{2}b_{1}c\right\} $,
$\left\{ a_{1}b_{2}c\right\} $ and $\left\{ a_{2}b_{2}c\right\} $.

Various methods have been discussed for evaluating complex event patterns.
One popular approach is to employ Non-deterministic Finite Automata
(NFAs) \cite{CugolaM10,DemersGHRW06,GyllstromADI08,PietzuchSB03,WuDR06}.
A query is compiled into an NFA consisting of a set of states and
conditional transitions between them. States are arranged in a pattern-specific
topology. Transitions are triggered by the arrival of an appropriate
event on the stream. Other mechanisms have also been proposed, such
as trees \cite{MeiM09}, finite state machines \cite{Schultz-MollerMP09},
and detection graphs \cite{AkdereMCT08}, to name a few.

An NFA for Example 1.1 is displayed in Figure 1. At each point in
time, an instance of the state machine is maintained for every detected
sub-match of the pattern. This instance is kept until a full match
is detected. As an example, consider again a stream $a_{1},a_{2},b_{1},b_{2},c$.
After the first three events have arrived, the system will maintain
match prefixes $\left\{ a_{1}\right\} $, $\left\{ a_{2}\right\} $,
$\left\{ a_{1}b_{1}\right\} $ and $\left\{ a_{2}b_{1}\right\} $.
After the fourth event, two more instances will be added. Finally,
following the arrival of \textit{c}, the NFA detects four sequences
matching the pattern.

\begin{figure*}   \centering \includegraphics[scale=0.4]{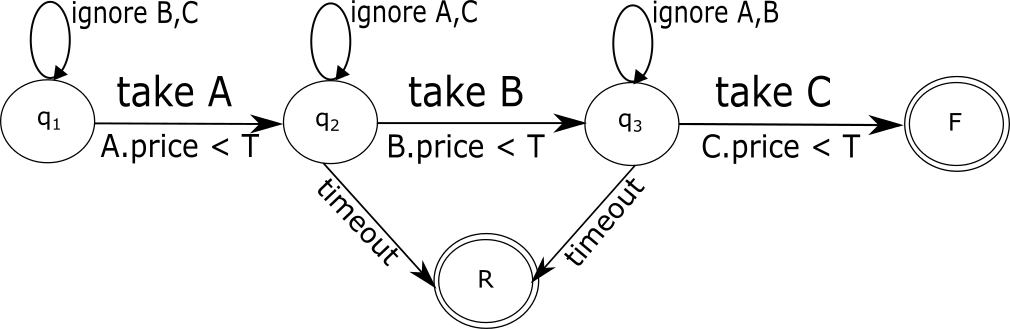}   \caption{NFA for the pattern from Example 1.1} \label{fig:NFA-for-Example} \end{figure*}

This approach, however, can prove inefficient when events arrive at
highly varied rates. Consider an input stream in which events of types
\textit{A} and \textit{B} arrive at a rate of one per second, whereas
an event of type \textit{C} arrives once every 12 hours. In this case,
the system must maintain a large number of prefixes that might not
lead to any matches. Since the number of prefixes to be kept grows
exponentially with the number of event types in a pattern, this method
becomes highly wasteful in terms of memory and computational resources.
The described situation could be avoided if the evaluation started
from \textit{C}, the rarest event type.

In this paper we propose a new NFA-based matching mechanism that overcomes
the above drawback. The proposed mechanism constructs and extends
partial matches by adding events in ascending order of frequency,
rather than according to their original order in the pattern. This
not only minimizes the number of partial matches held in memory, but
also reduces computation time, since there are fewer partial matches
to extend when processing a given event. Our proposed solution relies
on a lazy evaluation mechanism that can either process an event upon
arrival or store it in a buffer, referred to as the \emph{input buffer,}
to be processed at a later time if necessary.

In addition, we present two new NFA topologies that make use of the
lazy evaluation model to detect patterns; we call these topologies
a Chain NFA and a Tree NFA. A Chain NFA requires specifying the frequency
order of the event types. For example, to construct an automaton for
detecting the sequence $a,b,c$, it is necessary to specify that \textit{b}
is expected to be the most frequent event, followed by \textit{a,}
which is expected to be less frequent, followed by \textit{c,} which
is expected to be the least frequent. A Tree NFA also employs lazy
evaluation, but it \emph{does not }require specifying the frequency
order. Instead, it computes the actual order at each step in an ad
hoc manner.

Our lazy evaluation model applies to all common types of patterns,
including sequences, conjunctions, negations, and iterations. For
some of them, we believe our work is the first to produce an efficient
automata-based solution. The construction of a lazy NFA for each of
the aforementioned pattern types is presented in detail. We also extend
this topology for use with composite patterns. This extension, which
we call a Lazy Multi-Chain NFA, is capable of detecting an arbitrary
composition of the operators above. The correctness of all construction
algorithms is formally proven.

We experimentally evaluate our mechanism on real-world stock trading
data. The results demonstrate performance gain of two orders of magnitude
over traditional NFA-based approaches, with significantly reduced
memory resource requirements. It is also shown that for every stream
of events, the Tree NFA is at least as efficient as the best performing
Chain NFA.

The remainder of the paper is organized as follows. Section \ref{sec:Eager-Evaluation}
provides the required background and briefly describes the NFA evaluation
framework. In Section \ref{sec:Lazy-Evaluation}, the concepts and
ideas of lazy evaluation are presented. We proceed to describe how
a lazy Chain NFA can be constructed using given frequencies of the
participating events in Section \ref{sec:Lazy-Chain-NFA}. Lazy Multi-Chain
NFA is discussed in Section \ref{sec:Lazy-Multi-Chain-NFA}. We formally
prove the correctness of our construction in Section \ref{sec:Equivalence-of-the}.
We extend the above topologies to introduce a Tree NFA in Section
\ref{sec:Lazy-Tree-NFA}. Section \ref{sec:Experimental-Evaluation}
presents the experimental evaluation. Section \ref{sec:Related-Work}
describes related work. Section \ref{sec:Conclusions} summarizes
the paper.

\section{Eager NFA Evaluation}

\label{sec:Eager-Evaluation}

In this section, we discuss in detail the two main parts of a CEP
system: the specification language and the evaluation mechanism. For
the former, SASE+\cite{GyllstromADI08} will be assumed for the rest
of the paper. For the latter, we first present the ``eager'' NFA
evaluation framework over SASE+ patterns as described in \cite{AgrawalDGI2008},
processing every incoming event upon arrival. Then, the modifications
required to employ the lazy evaluation principle are described. The
SASE+ framework was chosen without loss of generality, and the same
ideas may be applied to any NFA-based method.

\subsection{Specification Language}

\label{sub:Specification-Language}

The SASE+ language, thoroughly described in \cite{GyllstromADI08},
combines a simple, SQL-like syntax with a high degree of expressiveness.
The semantics and expressive power of the language are precisely defined
in a formal model.

Each primitive event in SASE+ has an arrival timestamp, a type, and
a set of attributes associated with the type. An attribute is a data
item related to a given event type, represented by a name and a value.

In its most basic form, a complex event definition in SASE+ is composed
of three building blocks: PATTERN, WHERE and WITHIN. The PATTERN clause
defines the primitive events we would like to detect and the operators
applied to combine them into a pattern match. The WHERE clause specifies
constraints on the values of data attributes of the primitive events
participating in the pattern. These constraints may be combined using
Boolean expressions. Finally, the WITHIN clause defines a time window
over the entire pattern, specifying the maximal allowed time interval
between the arrivals of the primitive events.

As an example, consider the pattern from Example 1.1. One possible
representation of this complex event is:

\begin{equation} 
\begin{array}{l}
PATTERN\: SEQ\left(A\: a,B\: b,C\: c\right)\\ 
WHERE\: skip\_till\_any\_match\:\{\\ 
\qquad a.price>T\\ 
\qquad AND\: b.price>T\\ 
\qquad AND\: c.price>T\}\\ 
WITHIN\:1\: hour. 
\end{array} 
\end{equation} 

Here, we define three event types for stocks with identifiers \textit{A},
\textit{B} and \textit{C}. Every primitive event represents a value
of a respective stock at some point in time. Since a fixed order on
stock price reports is defined, the sequence (SEQ) operator is to
be used. We assume that a numerical attribute \textit{price} is assigned
to each event.\textit{ skip\_till\_any\_match} denotes the event selection
strategy as presented in Section \ref{sec:Introduction}.

A wide variety of operators is supported by SASE+. The most commonly
used operator types are sequence (SEQ), conjunction (AND), disjunction
(OR), negation (marked as '\textasciitilde{}'), and iteration (marked
as '+'). We will describe and discuss each of the operators in detail
in Section \ref{sec:Lazy-Chain-NFA}.

\subsection{The Eager Evaluation Mechanism}

\label{sub:The-Eager-Evaluation}

After a SASE+ pattern is created, it is compiled into an NFA, which
is then employed on an input stream to detect pattern occurrences.
In this section we define the structure of the eager NFA, which is
a slightly modified version of the one described in \cite{AgrawalDGI2008}.

Formally, an NFA is defined as follows:
\[
A=\left(Q,E,q_{1},F,R\right),
\]
where $Q$ is a set of states; $E$ is a set of directed edges; $q_{1}$
is an initial state; $F$ is an accepting state; and $R$ is a rejecting
state.

Evaluation starts at the initial state. Transitions between states
are triggered by event arrivals from the input stream. Each NFA instance
is associated with a \textit{match buffer}. As we proceed through
an automaton towards the final state, we use the match buffer to store
the events that caused the transitions. It is always empty at $q_{1}$,
and events are gradually added to it during evaluation. This is done
by executing actions of the traversed edges, as will be described
shortly.

The match buffer should be thought of as a logical construct. As discussed
in \cite{AgrawalDGI2008}, there is no need to allocate dedicated
memory for each match buffer, since multiple match buffers can be
stored in a compact manner that takes into account that certain events
may be included in many buffers.

If during the traversal of an NFA instance the accepting state is
reached, the content of the associated match buffer is returned as
a successful match for the pattern. If the rejecting state is reached,
the NFA instance and its match buffer are discarded. If the time window
specified in the WITHIN block is violated, a special \textit{timeout}
event is generated, resulting also in the rejecting state.

An edge is defined by the following tuple:
\[
e=\left(q_{s},q_{d},action,types,condition\right),
\]
where $q_{s}$ is the source state of an edge; $q_{d}$ is the destination
state; \textit{action} is always one of those described below; \textit{type}
may be one or more event types specified in the PATTERN block; and
\textit{condition} is a Boolean predicate, reflecting the conditions
in the WHERE block that have to be satisfied for the transition to
occur.

The \textit{action} associated with an edge is performed when the
edge is traversed. It can be one of the following (the actions listed
below are simplified versions of the ones defined for SASE \cite{AgrawalDGI2008}):
\begin{itemize}
\item \textit{take} - consumes the event from the input stream and adds
it to the match buffer. A new instance is created.
\item \textit{ignore} - skips the event (consumes the event and discards
it instead of storing). No new instance is created.
\end{itemize}
As an example, consider again Figure \ref{fig:NFA-for-Example}. Note
that the accepting state can only be reached by executing three \textit{take}
actions; hence, successful evaluation will produce a match buffer
containing three primitive events which comprise the detected match.

\begin{figure*}   \includegraphics[width=\textwidth]{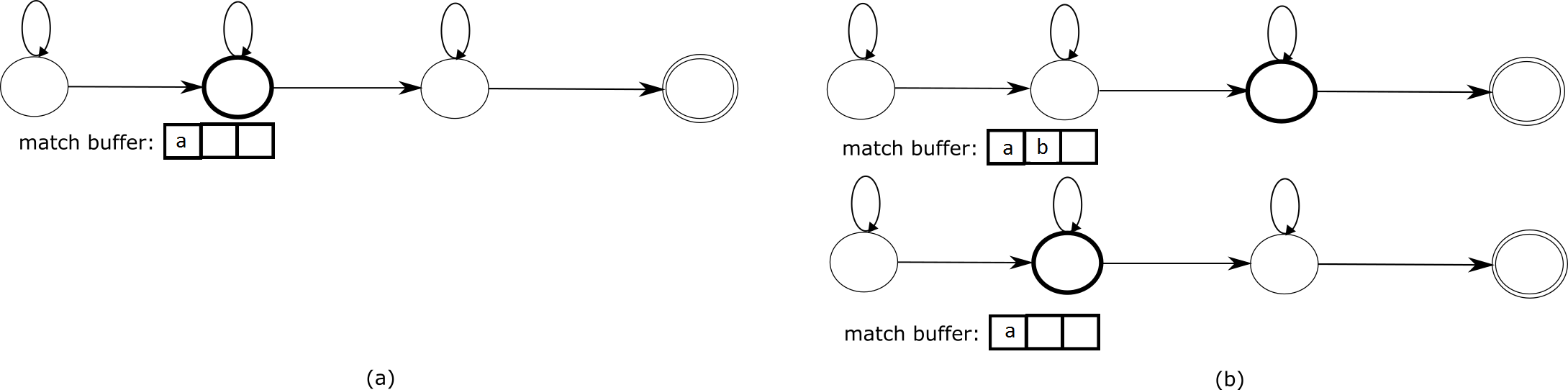}   \caption{Non-deterministic evaluation of NFA for Example 1.1. (a) The sole NFA instance is currently at the second evaluation stage, with a single event in its match buffer. (b) A new event g arrives, and now the NFA instance can either (1) accept the new event as a part of the potential match and proceed to the next step, or (2) ignore it (by traversing a self loop) and keep waiting for a future event of the same type. The problem is solved by duplicating the instance and applying both moves.} \label{fig:Non-determistic-evaluation-of} \end{figure*}

Several edges may lead from the same state and specify the same event
type. In this case, an event will non-deterministically cause multiple
traversals from a given state. If an event triggered \textit{n} edges,
the instance will be replicated \textit{n} times, for each of the
possible traversals. As an example, consider the situation described
in Figure \ref{fig:Non-determistic-evaluation-of} (for simplicity,
the rejecting state is omitted). In \ref{fig:Non-determistic-evaluation-of}(a),
there is some instance of an NFA from Figure \ref{fig:NFA-for-Example}
with an event $a$ in its match buffer, currently in state $q_{2}$
(we mark the current state of an instance with bold border). In \ref{fig:Non-determistic-evaluation-of}(b),
an event $b$ has arrived. This event triggers the traversal of two
edges, namely the outgoing \textit{take} edge and the outgoing \textit{ignore}
edge. As a result, one new instance will be created to allow both
traversals to occur.

Upon startup, the system creates a single instance with an empty match
buffer, whose current state is the initial state. As events arrive,
multiple instances of an NFA are created and run in parallel, one
for each partial match detected up to that point. Every event received
on the input stream will be applied to all NFA instances.

\begin{figure}   \centering \includegraphics[width=.8\linewidth]{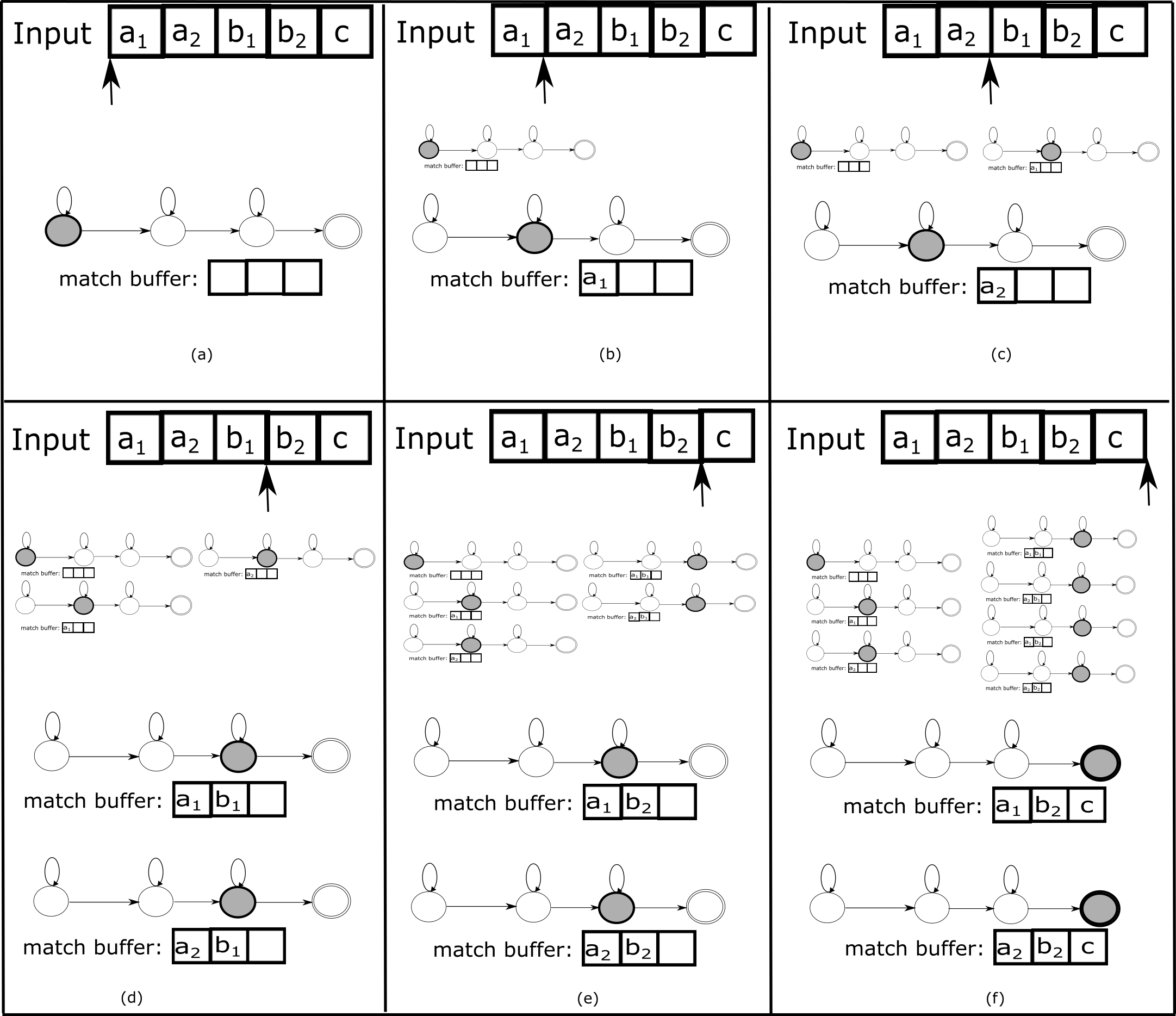}   \caption{Example of a non-deterministic evaluation of the NFA from Figure \ref{fig:NFA-for-Example}. The rejecting state is omitted for simplicity. At each step, newly created instances are magnified. The current state of each instance is highlighted in gray.} \label{fig:Example-of-Eager} \end{figure}

Figure \ref{fig:Example-of-Eager} illustrates the eager evaluation
process of the NFA from Figure 1, applied on an input stream $a_{1},a_{2},b_{1},b_{2},c$.

\section{Lazy NFA Evaluation}

\label{sec:Lazy-Evaluation}

In this section, we describe the principles of the lazy evaluation
method and the steps required to implement it over an eager NFA framework.
Specific NFA topologies applying the lazy evaluation paradigm on different
types of patterns will be discussed in the subsequent sections.

First, we will exemplify the need for such a mechanism. As demonstrated
in Section \ref{sec:Introduction}, more frequent primitive events
might trigger creation of instances that do not lead to any matches.
Consider, for example, how the evaluation in Figure \ref{fig:Example-of-Eager}
might look if the input stream contained 100 events of type \textit{A},
followed by 100 events of type \textit{B}. Since no event of type
\textit{C} is present, no pattern match can exist. However, $100^{2}$
instances would be spawned and redundant computations would take place.
This could have been avoided simply by modifying the evaluation order
to start from \textit{C}.

There are two main reasons for the problem described above. First,
an eager NFA is constructed in a manner that preserves the structure
of the underlying pattern. Second, any primitive event must be processed
immediately upon arrival. Hence, e.g., in a sequence pattern, re-ordering
events with mutual temporal constraints is not possible.

The lazy evaluation model takes advantage of the varying arrival rates
of the events in the pattern to significantly reduce the use of computational
and memory resources. It can be easily implemented by applying a number
of modifications on the eager NFA model, which we will present shortly.
We assume that all frequencies are known in advance (this assumption
will be removed in Section \ref{sec:Lazy-Tree-NFA}). 

The idea behind lazy evaluation is to enable instances to store incoming
events, and if necessary, retrieve them later for processing. This
functionality allows us to process incoming events in any arbitrary
order, rather than in the order of their appearance in the input stream.
Specifically, it can be utilized to arrange NFA states according to
the order of event frequencies, as we will exemplify in Section \ref{sec:Lazy-Chain-NFA}.

To achieve out-of-order event processing, an additional buffer, referred
to as the \textit{input buffer}, is associated with each NFA instance,
and an additional edge action, referred to as \emph{store,} is defined.
When an edge with a \textit{store} action is traversed, the event
causing the traversal is inserted into the input buffer. The input
buffer stores events in chronological order.

Events located in the input buffer can later be fetched for processing.
This functionality is implemented by extending the semantics of the
\textit{take} action. Whereas in the eager NFA model an event accepted
by this type of edge is always taken from the input stream, here it
also triggers a search for events of the required type inside the
input buffer. The results of this search, as well as events from the
input stream, are then evaluated non-deterministically by spawning
new NFA instances. If the search yields no result, or if the retrieved
events fail to trigger a \textit{take} transition due to unsatisfied
conditions, a special event called \textit{search\_failed} is generated
for the current instance.

\begin{figure}   \centering \includegraphics[scale=0.4]{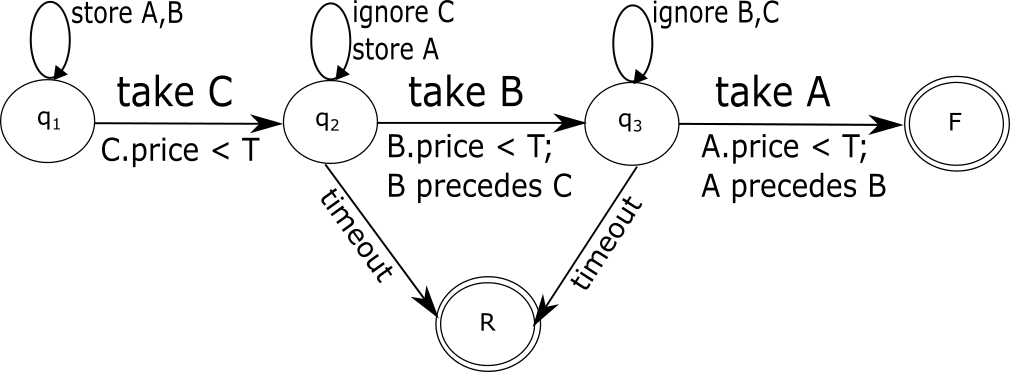}   \caption{Lazy Chain NFA for Example 1.1 with frequency order \textit{C,B,A}.} \label{fig:Lazy-NFA-for-Example} \end{figure}

Consider again the pattern from Example 1.1. Assume that the event
type \textit{C} is the least frequent, and the event type \textit{A}
is the most common. Figure \ref{fig:Lazy-NFA-for-Example} shows a
possible lazy NFA, which detects the pattern in order \textit{C,B,A}.
Upon arrival of \textit{A} or \textit{B}, an outgoing \textit{store}
edge of $q_{1}$ inserts them into the input buffer. Following an
arrival of type \textit{C}, an outgoing \textit{take} edge of $q_{2}$
retrieves events belonging to type \textit{B} from the input buffer
and attempts to add each of them to a partial match, as if they arrived
from the input stream. The same occurs for events of type \textit{A}
in $q_{3}$.

Note that invoking a full scan of the entire input buffer on each
\textit{take} action would be inefficient and unnecessary. While searching
for events of type \textit{A}, we are only interested in those that
arrived before the already accepted \textit{B}. In general, only a
certain, usually very narrow range of events in the input buffer is
relevant to a given \textit{take} edge. Since the input buffer is
arranged chronologically, these temporal constraints make it possible
to efficiently limit the search to only a small portion of the storage.

To implement the desired functionality, we extend the NFA edge definition:
\[
e=\left(q_{s},q_{d},action,types,condition,prec,succ\right).
\]

Here, \textit{prec} and \textit{succ}, also known as \textit{ordering
filters}, are sets of event types which enforce temporal limitations
on an event taken by this edge. Elements in \textit{prec} must precede
this event, while elements in \textit{succ} must succeed it. Both
sets may only contain event types selected among those already taken
during evaluation until this point. In the example above, for the
edge taking \textit{A,} the following holds: $prec\left(A\right)=\varnothing;succ\left(A\right)=\left\{ B,C\right\} $.
Thus, while attempting to traverse this edge, the system will scan
the buffer from the beginning, but only until the timestamp of \textit{B}
is reached. For an event type \textit{E} to be accepted from the input
stream, we require the condition $succ\left(E\right)=\varnothing$.

The following sections will provide formal definitions of the different
types of lazy NFAs, including the corresponding ordering filter definitions.

\section{Lazy Chain NFA}

\label{sec:Lazy-Chain-NFA}

The Chain NFA is the first of the two proposed NFA topologies utilizing
the constructs of the lazy evaluation model presented in Section \ref{sec:Lazy-Evaluation}.
This section presents the universal, pattern-independent definition
of the Chain NFA and its applications for detecting a wide range of
pattern types. The second topology, which we refer to as the Tree
NFA, is discussed in Section \ref{sec:Lazy-Tree-NFA}. 

We will start by providing the intuition behind the construction of
a generic lazy Chain NFA. All pattern types discussed in this section
will share this common structure with only minor adaptations. The
following sub-sections will focus on each type, providing formal and
detailed definitions.

Given a pattern containing $n$ primitive event types, a corresponding
Chain NFA consists of $n+2$ states. The first $n+1$ states are arranged
in a chain according to ascending frequency order of the events, which
we assume to be given in advance. Each of the first $n$ states is
responsible for detecting one primitive event in the pattern (the
initial state detects the rarest event, the next state detects the
second rarest event, etc.). The final state in a chain is the accepting
state \textit{F}. In addition, the rejecting state \textit{R} is connected
to all states except for $q_{1}$ and \textit{F}. Its purpose is to
collect invalid and expired partial matches.

We will denote by \textit{freq} the ascending frequency order in which
the first $n$ states are arranged. We will also denote by $e_{i}$
the $i^{th}$ event type in \textit{freq} and by $q_{i}$ the corresponding
state in the chain. The state $q_{i}$ will generally have several
types of outgoing edges. A \textit{take} edge attempts to add the
next event in the pattern to the partial match and to advance to the
next state $q_{i+1}$. A \textit{store} edge adds all events of types
yet to be processed (succeeding $e_{i}$ in \textit{freq}) to the
input buffer. An \textit{ignore} edge discards events of already processed
types (preceding $e_{i}$ in \textit{freq}). Finally, a \textit{timeout}
edge accepts the special \textit{timeout} event and proceeds to \textit{R}.

Figure \ref{fig:Lazy-NFA-for-Example} exemplifies the common structure
of a Chain NFA. Note that it is based solely on relative frequencies
of primitive events. It depends neither on the operator(s) applied
by the input pattern nor on its structure (e.g., on the requested
temporal sequence order). Instead, pattern-specific requirements will
be expressed by the parameters of the edges, as will be explained
below.

\subsection{Sequences}

\label{sub:Sequence}

Sequences are patterns requiring a number of primitive events to arrive
in a predefined order. Pattern 1 presented in Section \ref{sub:Specification-Language}
is an example of a sequence of three events. A Chain NFA for sequence
patterns conforms fully to the common structure. Since temporal constraints
are crucial for this pattern type, properly defining ordering filters
on edges is of particular importance.

We will proceed now to the formal description. Let $E_{i}$ denote
the set of outgoing edges of $q_{i}$. Let $Prec_{ord}\left(e\right)$
denote all events preceding an event $e$ in an order $ord$. Similarly,
let $Succ_{ord}\left(e\right)$ denote all events succeeding $e$
in $ord$. Then, $E_{i}$ will contain the following edges:
\begin{itemize}
\item $e_{i}^{ignore}=\left(q_{i},q_{i},ignore,Prec_{sel}\left(e_{i}\right),true,\varnothing,\varnothing\right)$:
any event whose type corresponds to one of the already taken events
is ignored.
\item $e_{i}^{store}=\left(q_{i},q_{i},store,Succ_{sel}\left(e_{i}\right),true,\varnothing,\varnothing\right)$:
any event that might be taken in one of the following states is stored
in the input buffer.
\item $e_{i}^{timeout}=\left(q_{i},R,ignore,timeout,true,\varnothing,\varnothing\right)$:
if a timeout event is detected, the NFA instance proceeds to the rejecting
state and is subsequently discarded.
\item $e_{i}^{take}=\left(q_{i},q_{i+1},take,e_{i},cond_{i},prec_{i},succ_{i}\right)$:
an event of type $e_{i}$ is taken only if it satisfies the conditions
required by the initial pattern (denoted by $cond_{i}$).
\end{itemize}
Now we will define how ordering filters $prec_{i}$ and $succ_{i}$
are calculated. Given a set $S$ of events, let $Latest\left(S\right)$
be the latest event in $S$ (i.e., an event with the largest timestamp
value). Correspondingly, let $Earliest\left(S\right)$ denote the
earliest event in $S$. Finally, let $seq$ denote the original sequence
order as specified by the input pattern.

The ordering filters for a \textit{take} edge $e_{i}^{take}$ will
be defined as follows:

\[
prec_{i}=\begin{cases}
\begin{array}{c}
Latest\left(Prec_{freq}\left(e_{i}\right)\cap Prec_{seq}\left(e_{i}\right)\right)\\
\slashed{O}
\end{array} & \begin{array}{c}
if\; Prec_{freq}\left(e_{i}\right)\cap Prec_{seq}\left(e_{i}\right)\neq\textrm{Ø}\\
otherwise
\end{array}\end{cases};
\]

\[
succ_{i}=\begin{cases}
\begin{array}{c}
Earliest\left(Prec_{freq}\left(e_{i}\right)\cap Succ_{seq}\left(e_{i}\right)\right)\\
\slashed{O}
\end{array} & \begin{array}{c}
if\; Prec_{freq}\left(e_{i}\right)\cap Succ_{seq}\left(e_{i}\right)\neq\textrm{Ø}\\
otherwise
\end{array}.\end{cases}
\]

It can be seen that $prec_{i}$ and $succ_{i}$ consist of a single
element each. $prec_{i}$ contains the latest event preceding $e_{i}$
in the original sequence order, out of those already accepted when
the state $q_{i}$ is reached. Similarly, $succ_{i}$ holds the earliest
event, out of those available, succeeding $e_{i}$ in $seq$.

The Chain NFA for sequence patterns will thus be defined as follows:
\[
\begin{array}{c}
A=\left(Q,E,q_{1},F,R\right);\\
Q=\left\{ q_{i}|1\leq i\leq n\right\} \cup\left\{ F,R\right\} ;\; E=\bigcup_{i=1}^{n}E_{i}.
\end{array}
\]

Figure \ref{fig:Lazy-NFA-for-Example} demonstrates the Chain NFA
for the pattern from Section \ref{sub:Specification-Language}.

\subsection{Conjunctions}

\label{sub:Conjunction}

Conjunction patterns detect a set of events in the input stream, regardless
of their mutual order of arrival. This pattern type presents a considerable
challenge to the traditional NFA-based approaches. The complication
follows from the nature of a finite automaton, which requires specifying
an order in which pattern elements will be accepted. However, as all
orders between the participating events are valid, the only legitimate
way to construct an NFA for the desired pattern is by incorporating
all of them.

Consider the following simple conjunction pattern:

\begin{equation} 
\begin{array}{l} 
PATTERN\: AND\left(A\: a,B\: b,C\: c\right)\\ 
WITHIN\:1\: hour. 
\end{array} 
\end{equation}

\begin{figure}   \centering \includegraphics[width=.65\linewidth]{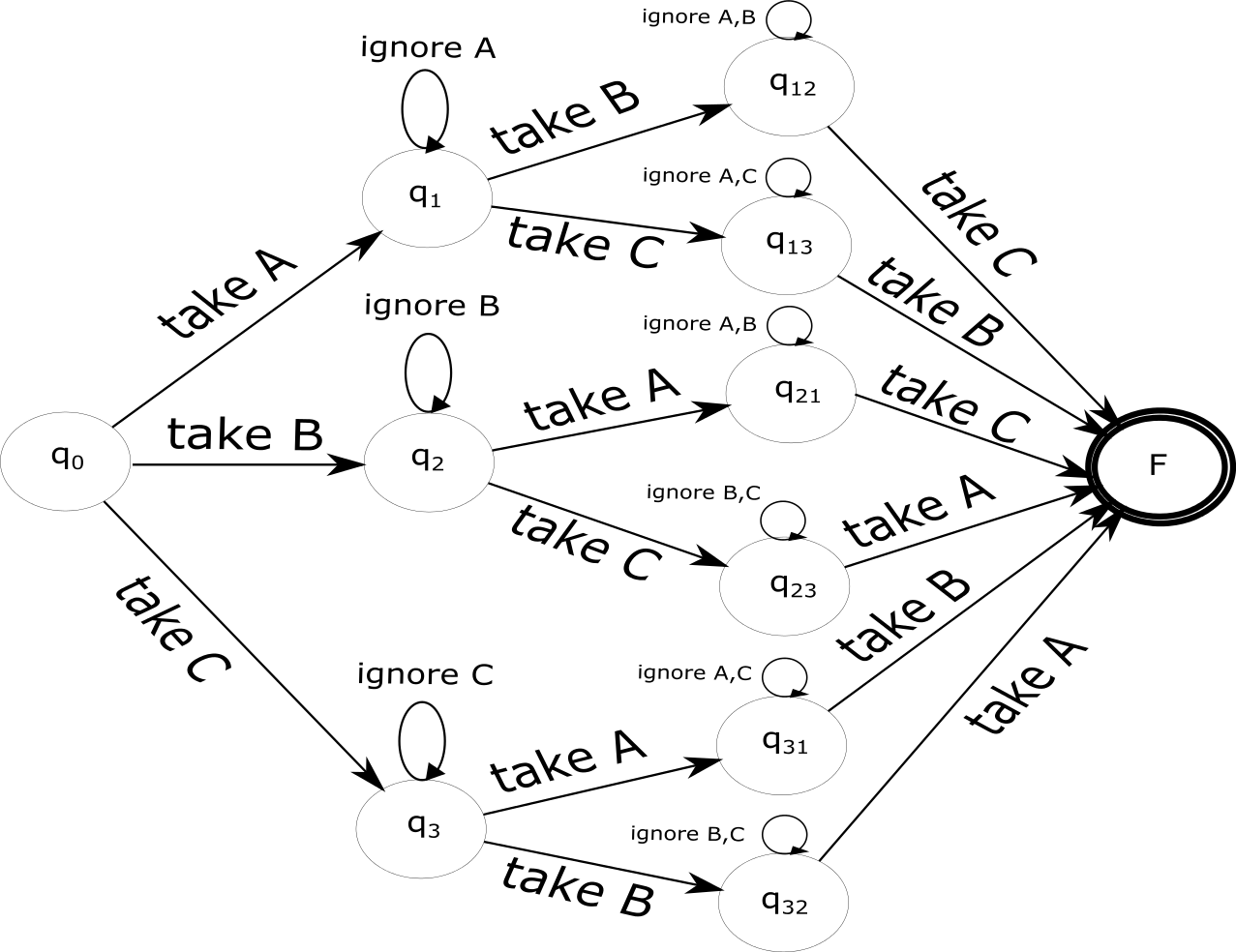}   \caption{Eager NFA for Conjunction Pattern \textit{AND(A,B,C)}. The rejecting state is omitted for simplicity.} \label{fig:Eager-NFA-for-Conjunction} \end{figure}Figure
\ref{fig:Eager-NFA-for-Conjunction} illustrates an eager NFA accepting
this pattern. It can be seen that the number of states and transitions
is exponential in length of the AND clause. As a result, the matching
process becomes highly inefficient, even for a small number of event
types.

The lazy Chain NFA solves the aforementioned problem. Instead of attempting
to match all possible orders, only the ascending frequency order is
incorporated, as presented in Figure \ref{fig:Lazy-NFA-for-Example}.
To the best of our knowledge, this is the first method for automata-based
detection of conjunction patterns that requires only a linear number
of states.

The definition of the Chain NFA for conjunctions is similar to the
one presented above for sequence patterns. The only difference is
the absence of ordering filters. Since no constraints can be defined
on the mutual order of primitive events, the entire content of the
input buffer has to be examined during each search operation, with
no possibility to discard parts of it. Thus, buffer searches are rather
slow as compared to searches that take place during evaluation of
sequences. However, this solution still significantly outperforms
the traditional (eager) approach, as our experimental results in Section
\ref{sec:Experimental-Evaluation} demonstrate.

More formally, a \textit{take} edge $e_{i}^{take}$ for an event type
$e_{i}$, detected by a state $q_{i}$, is defined as follows:
\[
e_{i}^{take}=\left(q_{i},q_{i+1},take,e_{i},cond_{i},,\varnothing,\varnothing\right).
\]

Otherwise, the construction is identical to the one shown in Section
\ref{sub:Sequence}.

\subsection{Partial Sequences}

\label{sub:Partial-Sequence}

Partial sequence patterns are conjunctions in which temporal constraints
exist between subsets of the primitive events involved. As an example,
consider a pattern:

\begin{equation} 
\begin{array}{l} 
PATTERN\: AND\left(SEQ\left(A\: a,B\: b\right),SEQ\left(C\: c,D\: d\right),E\: e\right)\\ 
WITHIN\:1\: hour. 
\end{array} 
\end{equation}

Here, an event of type \textit{B} must appear after an event of type
\textit{A}. However, it may arrive either before or after events of
types \textit{C}, \textit{D} and \textit{E}. Similarly, \textit{D}
has to appear after \textit{C}, and \textit{E} can appear at any place
in a match.

Note that sequence and conjunction patterns, as presented above, are
two opposite edge cases of a partial sequence.

The Chain NFA for a partial sequence will be defined identically to
the case of a full sequence. It will, however, incorporate ordering
filters only for those events participating in at least one sub-sequence.
Also, for events appearing in multiple sub-sequences, $prec_{i}$
and $succ_{i}$ will contain several values. Among those, only the
most restrictive ones will be chosen at runtime.

We will now proceed to the formal definition of ordering filters.
Let $SEQ=\left\{ seq_{1},\cdots,seq_{k}\right\} $ denote all sub-sequences
in a pattern. Note that the sub-sequences in \textit{SEQ} do not necessarily
contain independent sets. In addition, let 
\[
Prec_{SEQ}\left(e\right)=\bigcup_{seq\in SEQ}Prec_{seq}\left(e\right)
\]
\[
Succ_{SEQ}\left(e\right)=\bigcup_{seq\in SEQ}Succ_{seq}\left(e\right).
\]
The ordering filters for an edge $e_{i}^{take}$ are then defined
as follows:
\[
prec_{i}=\begin{cases}
\begin{array}{c}
Latest\left(Prec_{freq}\left(e_{i}\right)\cap Prec_{SEQ}\left(e_{i}\right)\right)\\
\slashed{O}
\end{array} & \begin{array}{c}
if\; Prec_{freq}\left(e_{i}\right)\cap Prec_{SEQ}\left(e_{i}\right)\neq\textrm{Ø}\\
otherwise
\end{array};\end{cases}
\]

\[
succ_{i}=\begin{cases}
\begin{array}{c}
Earliest\left(Prec_{freq}\left(e_{i}\right)\cap Succ_{SEQ}\left(e_{i}\right)\right)\\
\slashed{O}
\end{array} & \begin{array}{c}
if\; Prec_{freq}\left(e_{i}\right)\cap Succ_{SEQ}\left(e_{i}\right)\neq\textrm{Ø}\\
otherwise
\end{array}\end{cases}.
\]

Otherwise, the construction is identical to the one shown in Section
\ref{sub:Sequence}.

\subsection{Negations}

\label{sub:Negation}

In a pattern with negation, some of the primitive event types are
not allowed to appear at the predefined places. We will denote them
as \textit{negated events}. Negated events can be specified anywhere
in a pattern and form mutual conditions with positive events. Patterns
of any type may include a negation part.

The following is an example of a sequence pattern with a negated event:

\begin{equation} 
\begin{array}{l} 
PATTERN\: SEQ\left(A\: a,NOT\left(B\: b\right),C\: c,D\: d\right)\\ 
WHERE\: skip\_till\_any\_match\:\{b.x<c.y\}\\ 
WITHIN\:1\: hour. 
\end{array} 
\end{equation}

In this case, a successful pattern match will contain instances of
\textit{A}, \textit{C} and \textit{D} alone. Note that only events
of type \textit{B} satisfying the condition with an event of type
\textit{C} are not allowed to appear.

Existing NFA-based CEP frameworks employ two different techniques
for treating negated events. The first is to check for negative conditions
as a post-processing step, after the accepting state is reached. This
technique introduces a potential performance issue. Consider a case
in which events of type \textit{B} are very frequent and their presence
results in the discarding of all partial matches. Matches detected
by NFA will thus be invalidated only during post-processing, causing
superfluous computations. This situation could be avoided by moving
the negated event check to an earlier stage.

The second technique consists of augmenting an NFA with ``negative
edges'' that lead to a rejecting state upon detection of a forbidden
event. To the best of our knowledge, existing solutions of this kind
only solve limited cases. Namely, only sequences are considered and
no conditions between primitive events are supported. The reason for
these restrictions is that, when no event buffering is used, it is
impossible to verify the absence of a negated event that depends on
some future event. Consider again the example above. When an event
of type \textit{B} arrives, an eager NFA cannot check whether it satisfies
a condition with some event of type \textit{C}, since \textit{C} has
not yet been received. Our framework avoids this issue, as evaluation
is performed out of order.

We propose two Chain NFA-based solutions for detecting negations.
The first, which we call a Post-Processing Chain NFA, implements the
post-processing paradigm outlined above. The second, First-Chance
Chain NFA, attempts to detect a negated event as soon as possible.
From an analytical standpoint, neither solution is superior to the
other, and each can be favorable in different situations. A more efficient
mechanism for a particular negation pattern can be easily selected,
either automatically or manually.

\subsubsection{Post-Processing Negation}

\label{sub:Post-Processing-Negation}

The Lazy Post-Processing Chain NFA first detects a sub-chain of positive
events, and then proceeds to a second sub-chain, where each state
corresponds to a single negated event. For this negative sub-chain,
descending frequency order is used (as opposed to ascending order
for a positive part). Each \textit{negative state} is responsible
for verifying absence of one negated event. Thus, transitions between
negative states are triggered by either reaching a timeout or by an
unsuccessful search in the input buffer. These situations are indicated
by special \textit{timeout} and \textit{search\_failed} events. Arrival
of a forbidden event in a negative state triggers a transition to
the rejecting state. The last negative state is followed by the accepting
state.

\begin{figure}   \includegraphics[width=\linewidth]{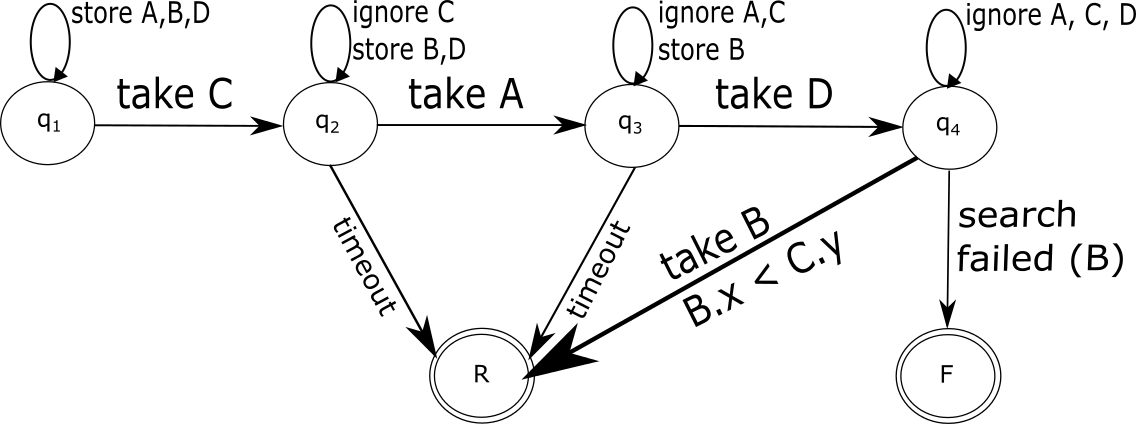}   \caption{Lazy Post-Processing Chain NFA for Pattern 4. The frequency order of \textit{C,A,D} is assumed. Ordering filters are omitted for clarity.} \label{fig:Post-Processing-Negation-NFA} \end{figure}

Figure \ref{fig:Post-Processing-Negation-NFA} demonstrates the Post-Processing
Chain NFA for Pattern 4. Since we only need to check for an occurrence
of \textit{B} before \textit{C}, a scan of the input buffer is sufficient.
Hence, only a \textit{search\_failed} event is expected. For patterns
in which a negated event may appear at the end (e.g., for conjunctions),
an edge taking \textit{timeout} is also required.

We will now formally define the Lazy Post-Processing Chain NFA.

Let $P=\left\{ e_{1},\cdots,e_{k}\right\} $ be all positive event
types in a pattern. Let $N=\left\{ h_{1},\cdots,h_{l}\right\} $ be
all negated event types. We will denote by $freq_{p}$ the ascending
frequency order of the events in \textit{P}, and by $freq{}_{n}$
the descending frequency order of the events in \textit{N}. Let $Q_{p}=\left\{ q_{1},\cdots,q_{k}\right\} $
be a set of states corresponding to positive events ordered according
to $freq_{p}$, and let $Q_{n}=\left\{ r_{1},\cdots,r_{l}\right\} $
be a set of states corresponding to positive events ordered according
to $freq{}_{n}$. Finally, let $E_{q}$ denote the set of outgoing
edges of a state \textit{q}.

For each positive state $q_{i}\in Q_{p};i\leq k$, the edges in $E_{q_{i}}$
are defined as for the Chain NFA for the underlying pattern, with
the exception of an edge $e_{i}^{store}$, which stores all events
in \textit{N} in addition to those in $Succ_{freq_{p}}\left(e_{i}\right)$.
Also, for $q_{k}$, the \textit{take} edge $e_{i}^{take}$ leads to
the first negative state, $r_{1}$.

For a negative state $r_{i}\in Q_{n};i\leq l$, the edges in $E_{r_{i}}$
are:
\begin{itemize}
\item $e_{i}^{ignore}=\left(r_{i},r_{i},ignore,Prec_{freq_{n}}\left(e_{i}\right)\cup P,true,\varnothing,\varnothing\right)$:
a positive event or previously checked negative event is ignored.
\item $e_{i}^{store}=\left(r_{i},r_{i},store,Succ_{freq_{n}}\left(e_{i}\right),true,\varnothing,\varnothing\right)$:
an event which may be potentially taken in one of the following states
is stored in the input buffer.
\item $e_{i}^{take}=\left(r_{i},R,take,h_{i},cond_{i},prec_{i},succ_{i}\right)$:
an instance of a negated event $h_{i}$ satisfying the conditions
triggers a transition to the rejecting state.
\end{itemize}
If $h_{i}$ can only be accepted from the input buffer (i.e., the
condition $succ\left(h_{i}\right)\neq\varnothing$ holds), an additional
edge is present:
\begin{itemize}
\item $e_{i}^{search\_failed}=\left(r_{i},r_{i+1},ignore,search\_failed,true,\varnothing,\varnothing\right)$:
in case of a failed search for a buffered event, the execution successfully
proceeds to the next state.
\end{itemize}
Otherwise, if $h_{i}$ can be accepted from the input stream, an additional
edge is present:
\begin{itemize}
\item $e_{i}^{timeout}=\left(r_{i},r_{i+1},ignore,timeout,true,\varnothing,\varnothing\right)$:
in case of a timeout, the negation test is passed and the execution
proceeds to the next state.
\end{itemize}
Also, for $r_{l}$, the \textit{timeout} or \textit{search\_failed}
edge leads to the accepting state $F$. The ordering filters $prec_{i}$
and $succ_{i}$ are calculated in the same manner as for the underlying
pattern type.

We are now ready to define the Lazy Post-Processing Chain NFA:
\[
\begin{array}{c}
A=\left(Q,E,q_{1},F,R\right);\\
Q=Q_{p}\cup Q_{n}\cup\left\{ F,R\right\} ;\; E=\bigcup_{q\in Q_{p}\cup Q_{n}}E_{q}.
\end{array}
\]

Although this NFA shares the previously discussed drawbacks of the
post-processing method, it also has several benefits over the other
negation NFA, described below. First, since event buffering is an
inherent part of the Lazy Evaluation mechanism, implementing this
NFA on top of the existing framework is straightforward. Second, in
some scenarios the best strategy is to postpone negation until the
end. One example is a very frequent negated event with a highly selective
filter condition. Third, this is the only possible approach for conjunctions
with negation and for sequences with a negated event at the end. In
these cases, we have to wait until the time window expires to perform
a negation check.

\subsubsection{First-Chance Negation}

\label{sub:First-Chance-Negation}

The Lazy First-Chance Chain NFA implements a paradigm opposite to
that of the Post-Processing Negation NFA. It operates by pushing the
detection of negated events to the earliest point possible. The key
observation is that it is often unnecessary to wait for all positive
events to arrive before launching a negated event check.

Consider Pattern 4 again. Clearly, a potential event of type \textit{B}
is only dependent on events of types \textit{A} (temporal condition)
and \textit{C} (explicit and temporal conditions). Consequently, we
only need to have \textit{A} and \textit{C} in our match buffer to
check for conflicting \textit{B} events. However, a solution based
on post-processing will also wait for \textit{D} to arrive. As a result,
if an event of type \textit{B} is found, it will lead to discarded
matches and to redundant operations, including the same search for
\textit{B} for every instance of \textit{D}. Furthermore, if \textit{D}
had mutual conditions with \textit{A} or \textit{C}, they would be
verified repeatedly, only to be invalidated later.

Lazy First-Chance Chain NFA overcomes this performance issue. For
each negated event, it determines the earliest state in which a check
for this event can be executed. From this state, a \textit{take} edge
is added, which leads to the rejecting state if an event matching
the conditions is encountered. This type of NFA therefore does not
have a state corresponding to each negated event. Instead, it consists
only of the positive chain, augmented with the aforementioned \textit{take}
edges.

As mentioned above, this approach cannot be efficiently applied on
a pattern in which a negated event may appear after all positive events
in a match. This is because for such an event, the earliest state
in which it can be detected is the accepting state.

\begin{figure}   \includegraphics[width=\linewidth]{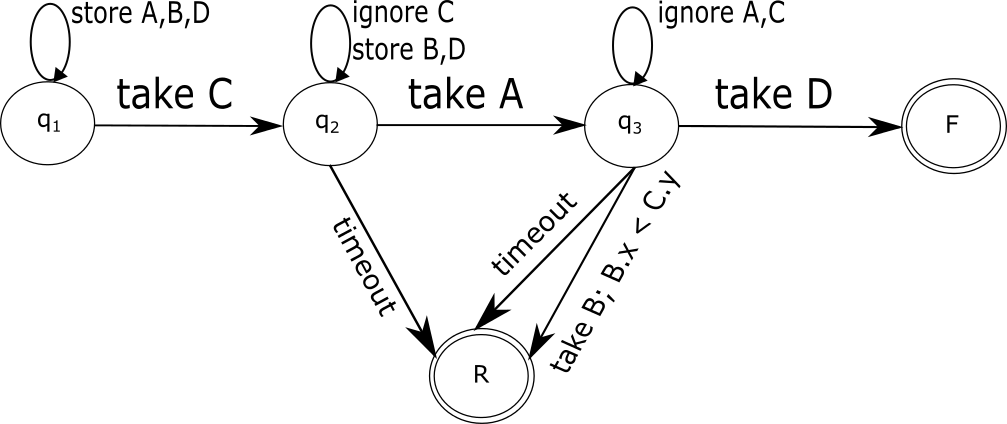}   \caption{Lazy First-Chance Chain NFA for Pattern 4. The frequency order of \textit{C,A,D} is assumed. For simplicity, ordering filters are omitted.} \label{fig:First-Chance-Negation-NFA} \end{figure}

Figure \ref{fig:First-Chance-Negation-NFA} demonstrates the First-Chance
Chain NFA for Pattern 4. As the figure shows, the absence of \textit{B}
is verified before \textit{D} is accepted, which is indeed the earliest
point possible for this pattern. Note that there are two edges between
$q_{3}$ and \textit{F}, one handling the timeout case and the other
detecting the negated event type \textit{B}.

We will now proceed to formally define the First-Chance Chain NFA.

Let $P=\left\{ e_{1},\cdots,e_{k}\right\} $ be all positive event
types in a pattern and let $N=\left\{ h_{1},\cdots,h_{l}\right\} $
be all negated event types. Let $A_{pos}=\left(Q_{pos},E_{pos},q_{1},F,R\right)$
denote a Chain NFA for the positive part of the pattern. For each
negated event $h_{i}$ we will define the following:
\begin{itemize}
\item $ImmPrec\left(h_{i}\right)=Latest\left(Prec_{SEQ}\left(h_{i}\right)\right)$,
the latest detected event preceding $h_{i}$.
\item $ImmSucc\left(h_{i}\right)=Earliest\left(Succ_{SEQ}\left(h_{i}\right)\right)$,
the earliest detected event succeeding $h_{i}$.
\item $Cond\left(h_{i}\right)$, set of all event types forming mutual conditions
with $h_{i}$.
\item $DEP\left(h_{i}\right)=\left\{ ImmPrec\left(h_{i}\right),ImmSucc\left(h_{i}\right)\right\} \cup Cond\left(h_{i}\right)$,
set of all event types which must be detected before the absence of
$h_{i}$ can be validated.
\item $q_{DEP}\left(h_{i}\right)$, the earliest state in $Q_{pos}$ in
which all events in $DEP\left(h_{i}\right)$ are already detected.
\end{itemize}
The First-Chance Chain NFA will then be constructed by augmenting
$E_{pos}$ with an edge from $q_{DEP}\left(h_{i}\right)$ to \textit{R}
for each $h_{i}$:
\[
\begin{array}{c}
A=\left(Q_{pos},E_{pos}\cup E_{rej},q_{1},F,R\right);\\
E_{rej}=\left\{ \left(q_{DEP}\left(h_{i}\right),R,take,h_{i},cond_{i}\right)|h_{i}\in N\right\} .
\end{array}
\]

In addition, each \textit{store} edge in $E_{pos}$ will be modified
to apply to events in \textit{N} as well.

\subsection{Iterations}

\label{sub:Iteration}

The term iteration operator (also called Kleene closure) refers to
patterns in which given events are allowed to appear multiple and
unbounded numbers of times. Detecting iterations is particularly challenging
under the skip-till-any-match event selection strategy because of
the exponential number of output combinations \cite{ZhangDI2014}.
Our solution utilizes lazy evaluation principles to minimize the number
of NFA instances and the calculations performed during the detection
process.

For clarity of presentation, we will only discuss sequence patterns
with a single iterated event. The concepts presented below can be
easily extended to conjunction and partial sequence patterns and to
an arbitrary number of events under iteration.

Consider the following sequence pattern with an iterated event: 

\begin{equation} 
\begin{array}{l} 
PATTERN\: SEQ\left(A\: a,B+\: b[],C\: c\right)\\ 
WITHIN\:1\: hour. 
\end{array} 
\end{equation}

For a sample input stream $a,b_{1},b_{2},b_{3},c$, the expected output
will be: 
\[
ab_{1}c,ab_{2}c,ab_{3}c,ab_{1}b_{2}c,ab_{1}b_{3}c,ab_{2}b_{3}c,ab_{1}b_{2}b_{3}c.
\]
Our approach is to convert this pattern into a regular sequence $SEQ\left(A,B,C\right)$
and to address the subsets of $b_{1},b_{2},b_{3}$ as separate ``events''.
For example, the input stream above can be converted to: 
\[
a(b_{1})(b_{2})(b_{3})(b_{1}b_{2})(b_{1}b_{3})(b_{2}b_{3})(b_{1}b_{2}b_{3})c.
\]
In this case our artificial ``\textit{B}+'' event type will become
the most frequent regardless of the original frequency of \textit{B}.
Consequently, if we were to construct an ordinary Chain NFA for this
sequence, we would place the state responsible for detecting \textit{B+}
at the end, as this event type would be the last in the ascending
frequency order.

Following this principle, two modifications are required for constructing
an Iteration Chain NFA from a Sequence Chain NFA. First, an iterated
event type has to be placed at the end of the frequency order, regardless
of its actual arrival rate. Second, the \textit{take} edge corresponding
to this event is required to produce all subsets of its instances.

To implement the second modification, we introduce a new edge action
called \textit{iterate}. An \textit{iterate} edge operates similarly
to a \textit{take} edge, consuming an event and adding it to the match
buffer. However, as an \textit{iterate} edge traverses the input buffer,
it produces subsets of events belonging to the required type and returns
them all. For example, if an NFA with an \textit{iterate} edge for
type \textit{B} is applied on a stream above, an input buffer will
contain 3 \textit{B} events and the \textit{iterate} edge will return
7 subsets, thus detecting 7 matches for the pattern.

When an event can be taken from the input stream, an \textit{iterate}
edge will add the new event to the input buffer, and then will only
generate subsets including this event.

A Chain NFA for Pattern 5 looks identical to the one displayed in
Figure 3 for Pattern 1, with an event type \textit{B} pushed to the
end and its corresponding edge action changed to \textit{iterate}.

Now we are ready to formally define a Lazy Iteration Chain NFA. Let
our pattern be $P=SEQ\left(e_{1},\cdots,e_{k}*,\cdots,e_{n}\right)$
and let $freq=e_{i_{1}},\cdots,e_{k},\cdots,e_{i_{n}}$ denote the
ascending frequency order of the primitive event types. The desired
automaton will be created by the following steps:
\begin{enumerate}
\item Define $freq'=e_{i_{1}},\cdots,e_{i_{n}},e_{k}$ (an order identical
to \textit{freq} except for moving $e_{k}$ to the end).
\item Create a Chain NFA $A_{seq}$ for \newline $P'=SEQ\left(e_{1},\cdots,e_{k},\cdots,e_{n}\right)$
with respect to the order $freq'$.
\item Let $e_{k}^{take}=\left(q_{e_{k}},F,take,e_{k},cond_{e_{k}},prec_{k},succ_{k}\right)$
denote the \textit{take} edge for $e_{k}$ in $A_{seq}$. Define \newline$e_{k}^{iterate}=\left(q_{e_{k}},F,iterate,e_{k},cond_{e_{k}},prec_{k},succ_{k}\right)$.
\item Produce a new NFA $A_{iterate}$ by replacing $e_{k}^{take}$ in $A_{seq}$
with $e_{k}^{iterate}$.
\end{enumerate}
More formally, if $A_{seq}=\left(Q,E,q_{1},F,R\right)$, then:
\[
A_{iterate}=\left(Q,(E\setminus\left\{ e_{k}^{take}\right\} )\cup\left\{ e_{k}^{iterate}\right\} ,q_{1},F,R\right).
\]

\subsubsection{Aggregations}

\label{sub:Gen-Aggregation}

Aggregation functions (SUM. AVG, MIN, MAX, etc.) can easily be integrated
into the framework using the presented approach for iteration patterns.
Since the lazy evaluation mechanism inherently supports event buffering,
aggregation is performed in a straightforward manner by invoking a
desired function on the input buffer contents. Note that an aggregation
function can only be applied on events under an iteration operator.
Consider the following example:

\begin{equation} 
\begin{array}{l} 
PATTERN\: SEQ\left(A\: a,B+\: b[],C\: c\right)\\ 
WHERE\: skip\_till\_any\_match\:\{AVG\left(b\left[i\right].x\right)<c.y\}\\ 
WITHIN\:1\: hour. 
\end{array} 
\end{equation} 

The condition will be evaluated upon traversal of a corresponding
\textit{iterate} edge. For each fetched subset of the available \textit{B}
events, the edge action will apply an aggregate function on this subset
and validate the condition.

\subsubsection{Repetitions}

\label{sub:Repetitions}

A repetition operator is a bounded version of an iteration. In repetition
patterns, we require a primitive event to appear at least \textit{l}
and at most \textit{m} times. An iteration pattern is thus a repetition
pattern with \textit{l} set to one and \textit{m} set to infinity.

Support for repetitions can be trivially added by defining lower and
upper bounds on the size of a subset returned by an \textit{iterate}
edge.

\subsubsection{Group-By-Attribute Optimization}

\label{sub:Group-By-Attribute-Optimization}

In some cases, creating all subsets of events before any processing
takes place is not the most efficient strategy. For example, consider
the following iteration pattern:

\begin{equation} 
\begin{array}{l} 
PATTERN\: SEQ\left(A\: a,B+\: b[],C\: c\right)\\ 
WHERE\: skip\_till\_any\_match\:\{b\left[i\right].x=b\left[i-1\right].x\}\\ 
WITHIN\:1\: hour. 
\end{array} 
\end{equation} 

For this pattern, most subsets of instances of \textit{B} will not
be valid, since all subset members are required to share the same
value of the attribute \textit{x}. Therefore, the evaluation procedure
described above will perform poorly, even compared to an eager approach.

For scenarios like the one shown, our framework introduces an optimization
that allows a user to specify a \textit{group-by-attribute}. Instances
of the iterated event are then hashed in the input buffer by the value
of this attribute. Upon traversal of an iterate edge, the generated
subsets will only contain events sharing the same attribute value.

\section{Lazy Multi-Chain NFA}

\label{sec:Lazy-Multi-Chain-NFA}

Despite the flexibility of the Chain NFA, its simple structure is
insufficient for detecting more complex patterns, involving separate
sets of events and matching conditions. One notable example is a pattern
featuring a disjunction operator. Disjunction patterns consist of
several sub-patterns, of which only one needs to be detected for the
successful match to be reported. Consequently, an NFA detecting a
disjunction must include multiple paths from the initial to the final
state. However, a Chain NFA only contains a single path between $q_{1}$
and \textit{F}, which renders it unsuitable for processing patterns
of this type.

\begin{figure}   \includegraphics[width=\linewidth]{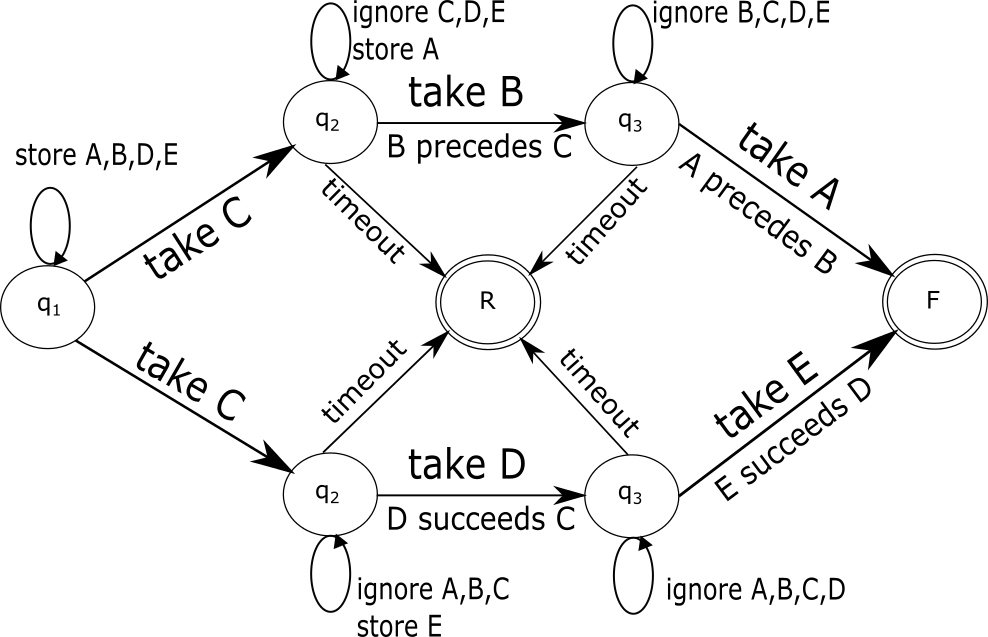}   \caption{Lazy Multi-Chain NFA for Pattern 8 with frequency order \textit{C,B,A,D,E}.} \label{fig:Lazy-Multi-Chain-NFA} \end{figure}

We will overcome this limitation by defining a new topology, which
we call \textit{Multi-Chain NFA}. This type of NFA will possess all
the qualities of the Chain NFA, but will also provide path selection
functionality. To construct it, we will first produce a Chain NFA
for each of the nested pattern parts. Then, we will merge the initial,
the accepting, and the rejecting states of all sub-automata and join
all the remaining states into a single automaton. The resulting NFA
will thus be structured as a union of chains leading to the accepting
state. Whenever a match for a sub-query is retrieved, one of the paths
will be traversed and the match will be reported. Figure \ref{fig:Lazy-Multi-Chain-NFA}
demonstrates an example of this construction.

In Section \ref{sub:Disjunction} we will formally describe the structure
of a Multi-Chain NFA for disjunction patterns. We will then present
a generalization to composite patterns in Section \ref{sub:Composite-Patterns}.

\subsection{Disjunctions}

\label{sub:Disjunction}

Disjunctions are the most commonly used type of composite patterns.
They consist of multiple sub-patterns, which can belong to either
of the types discussed above. As a set of events satisfying at least
a single sub-pattern is detected, it is reported as a match for the
whole pattern. For example, consider the following:

\begin{equation} 
\begin{array}{l} 
PATTERN\: OR\left(SEQ\left(A\: a,B\: b,C\: c\right),SEQ\left(C\: c,D\: d,E\: e\right)\right)\\ 
WITHIN\:1\: hour. 
\end{array} 
\end{equation} 

A Lazy Multi-Chain NFA for this pattern is displayed in Figure \ref{fig:Lazy-Multi-Chain-NFA}.
Note that a single primitive event is allowed to appear in multiple
sub-patterns. In this case, upon arrival of an event of type \textit{C},
two edge traversals from $q_{1}$ will be executed in a non-deterministic
manner and two new NFA instances will be created.

We will now proceed to the formal definition. Let $p_{1},\cdots,p_{m}$
be the sub-patterns of the disjunction pattern and let
\[
A_{1}=\left(Q_{1},E_{1},q_{1}^{1},F_{1},R_{1}\right)
\]
\[
\vdots
\]
\[
A_{m}=\left(Q_{m},E_{m},q_{1}^{m},F_{m},R_{m}\right)
\]
denote the Chain NFA for $p_{1},\cdots,p_{m}$.

Let $q_{1},F,R$ be the initial, the accepting, and the rejecting
states of the new Multi-Chain NFA $A_{OR}$, respectively.

The following definitions describe the outgoing edges of the initial
states of $A_{1},\cdots,A_{m}$ and the incoming edges of their final
(accepting and rejecting) states.
\[
E_{start}^{j}=\left\{ e|e=\left(q_{1}^{j},r,action,type,condition\right)\right\} ;
\]
\[
E_{acc}^{j}=\left\{ e|e=\left(q,F_{j},action,type,condition\right)\right\} ;
\]
\[
E_{rej}^{j}=\left\{ e|e=\left(q,R_{j},action,type,condition\right)\right\} ;
\]
\[
E_{start}=\bigcup_{j=1}^{m}E_{start}^{j};\: E_{acc}=\bigcup_{j=1}^{m}E_{acc}^{j};\: E_{rej}=\bigcup_{j=1}^{m}E_{rej}^{j}.
\]

Now, we will define the new edges that will replace the existing ones
and lead to $q_{1},F,R$ in the new NFA.
\[
E_{OR-start}^{j}=\left\{ e|e=\left(q_{1},r,action,type,condition\right)\right\} ;
\]
\[
E_{OR-acc}^{j}=\left\{ e|e=\left(q,F,action,type,condition\right)\right\} ;
\]
\[
E_{OR-rej}^{j}=\left\{ e|e=\left(q,R,action,type,condition\right)\right\} ;
\]
\[
E_{OR-start}=\bigcup_{j=1}^{m}E_{OR-start}^{j};
\]
\[
E_{OR-acc}=\bigcup_{j=1}^{m}E_{OR-acc}^{j};
\]
\[
E_{OR-rej}=\bigcup_{j=1}^{m}E_{OR-rej}^{j}.
\]

In addition, let
\[
Q_{start}=\left\{ q_{1}^{j}|1\leq j\leq m\right\} ;
\]
\[
Q_{F}=\left\{ F_{j}|1\leq j\leq m\right\} ;
\]
\[
Q_{R}=\left\{ R_{j}|1\leq j\leq m\right\} .
\]

Now, we are ready to define the Multi-Chain NFA:
\[
\begin{array}{c}
A_{OR}=\left(Q,E,q_{1},F,R\right);\\
Q=\left(\left(\bigcup_{j=1}^{m}Q_{j}\right)\setminus\left(Q_{start}\cup Q_{F}\cup Q_{R}\right)\right)\cup\left\{ q_{1},F,R\right\} ;\\
E=\left(\left(\bigcup_{j=1}^{m}E_{j}\right)\setminus\left(E_{start}\cup E_{acc}\cup E_{rej}\right)\right)\cup\\
\left(E_{OR-start}\cup E_{OR-acc}\cup E_{OR-rej}\right).
\end{array}
\]

\subsection{General Composite Patterns}

\label{sub:Composite-Patterns}

A disjunction pattern can be alternatively viewed as a Boolean formula
normalized to its DNF form. Moreover, all non-unary operators presented
earlier can be expressed as operations of Boolean calculus%
\footnote{For the purpose of the DNF conversion procedure, we represent a sequence
pattern (full or partial) as a conjunction with additional temporal
conditions between primitive events.%
}. Since any Boolean statement can be converted to DNF, it is possible
to employ a Lazy Multi-Chain NFA for matching an arbitrary composite
SASE+ pattern. The only step to be added to the construction algorithm
from the previous section is transforming the input pattern accordingly.

It should be noted, however, that applying the above procedure may
cause some sub-expressions to appear in multiple branches of the main
disjunction pattern. This will, consequently, lead to superfluous
calculations and NFA instances. For example, consider the following
pattern:

\begin{equation} 
\begin{array}{l} 
PATTERN\: SEQ\left(A\: a,OR\left(B\: b,C\: c\right),D\: d\right)\\ 
WHERE\: skip\_till\_any\_match\:\{a.price>d.price\}\\ 
WITHIN\:1\: hour. 
\end{array} 
\end{equation} 

The DNF form of this pattern is
\[
OR\left(SEQ\left(A\: a,B\: b,D\: d\right),SEQ\left(A\: a,C\: c,D\: d\right)\right).
\]

The derived Multi-Chain NFA will contain two branches, both of which
may perform computations for same pairs of events of types \textit{A}
and \textit{D}. The problem becomes even more severe if \textit{B}
and \textit{C} are negated types. In this case, the same output match
\textit{<a,d>} may be reported twice, since it satisfies both sub-patterns.

This issue may be solved by applying known multi-query techniques,
e.g., as described in \cite{DemersGHRW06}, and is a subject of our
future work.

\section{Equivalence of the Eager and the Lazy Chain NFA}

\label{sec:Equivalence-of-the}

In this section, we formally prove the correctness of the Chain NFA
construction. It is shown that, for any pattern over the presented
operators, a Chain NFA or a Lazy Multi-Chain NFA constructed according
to Section \ref{sec:Lazy-Chain-NFA} or \ref{sec:Lazy-Multi-Chain-NFA}
respectively is equivalent to the corresponding eager NFA in terms
of the language it accepts.

\newtheorem{theorem}{Theorem}
\begin{theorem}

Let $P$ be a complex event pattern, as defined in Section \ref{sub:Specification-Language},
over the event types $e_{1},\cdots,e_{n}$. Let $A_{eager}=(Q_{eager},E_{eager},q_{1}^{eager},F_{eager},R_{eager})$
be an eager NFA derived from $P$ as described in Section \ref{sub:The-Eager-Evaluation}
and let $A_{lazy}=(Q_{lazy},E_{lazy},q_{1}^{lazy},F_{lazy},R_{lazy})$
be a Chain NFA or a Multi-Chain NFA derived from $P$. Then $A_{eager}$
and $A_{lazy}$ are equivalent, i.e., $A_{eager}$ accepts a stream
of events if and only if $A_{lazy}$ accepts it.

\end{theorem}

We will start with the outline of the proof. It will proceed in several
steps.

First, we examine the case where $P$ is a pure sequence pattern.
We show that swapping two adjacent states in the NFA detecting $P$
does not affect its correctness when the input buffer is used. From
there, the statement is proven by induction for an arbitrary detection
order. Next, the case of a sequence with negation is shown separately
for both lazy mechanisms introduced above. In both parts, NFA construction
properties are used to infer correctness. We prove the case of a sequence
with iteration by induction on the length of an iterated pattern match,
viewing the pattern as a regular sequence during the induction step.

Afterwards, we continue to more complex pattern types. We start by
observing that conjunctions and partial sequences can be represented
as disjunctions of full sequences. Then, we prove that two branch-structured
NFAs (i.e., built of several chains starting at the initial state)
are equivalent if their sub-automata are equivalent. Similarly, we
demonstrate that a branch-structured NFA is equivalent to a chain-structured
NFA, if all of the branches are equivalent to this chain-structured
NFA. On the basis of these three statements, we derive the correctness
for conjunctions and partial sequences as well. From there, the case
of disjunctions is shown in a similar manner. We complete the proof
by generalizing this case for an arbitrary composite pattern, using
DNF form conversion.

\newtheorem{lemma}{Lemma}
\begin{lemma}

Let $P$ be a sequence pattern over $e_{1},\cdots,e_{n}$, let $seq$
be the order of the events in $P$, and let $freq$ be an ascending
frequency order. Let $A_{eager}$ be an eager NFA and let $A_{lazy}$
be a Chain NFA derived from $P$. Then $A_{eager}$ and $A_{lazy}$
are equivalent.

\end{lemma}

We first address the case where the order $freq$ is identical to
$seq$, i.e., we show that a lazy NFA for $P$ where the frequency
order is identical to the sequence order is equivalent to the eager
NFA for $P$. Next we show that if $freq'$ is an order received by
swapping the event types in location $i$ and $i+1$ ($1\leq i\leq n-1$)
in $freq$, then $A_{lazy}$ is equivalent to $A_{lazy'}$, which
uses $freq'$ instead of $freq$. This will conclude the proof, since
we can obtain $freq$ by a set of swaps of adjacent event types on
$seq$.

For the case $seq=freq$, the ordering filters of any edge will point
to the end of the input buffer, and all events will be taken from
the input stream; hence, the conditions on edges will become the same
as in the eager NFA. In addition, the order of states will be the
same as in the eager NFA. Consequently, the transition function between
states is the same, making the two automata identical.

Now assume that $freq'$ is received by swapping the event types at
location $i$ and $i+1$. Let $A_{lazy'}$ be a corresponding Chain
NFA derived from $P$ and $freq'$.

Note that $A_{lazy}$ and $A_{lazy'}$ differ only in the definitions
of the outbound edges (including the self-loops) from the $i^{th}$
and $\left(i+1\right)^{th}$ states, in particular in the ordering
filters on their $take$ edges. Let $s_{1},...,s_{m}$ be a stream
of events. We examine the processing of this stream by both $A_{lazy}$
and $A_{lazy'}$. We focus on instances of $A_{lazy}$ that have reached
the state $q_{i+2}^{lazy}$. We show that for every such instance,
there is a corresponding instance in $A_{lazy'}$ that has reached
$q_{i+2}^{lazy'}$ with the same input and match buffers. It is sufficient
to show one direction (that instances in $A_{lazy}$ have corresponding
instances in $A_{lazy'}$), since the proof for the other direction
is symmetric. Proving such a mapping will show that $A_{lazy}$ and
$A_{lazy'}$ are equivalent, since the processing of these instances
will be identical in both NFAs from this point on.

Let us examine an instance $I$ that has reached $q_{i+2}^{lazy}$
in $A_{lazy}.$ We will denote the event that caused $I$ to transition
into $q_{i+2}^{lazy}$ by $s_{l}$. In addition, let us denote the
event that caused $I$ to transition into $q_{i}^{lazy}$ by $s_{k}$.
Let $t_{k}$ denote the arrival time of $s_{k}$. Since the definitions
of $A_{lazy}$ and $A_{lazy'}$ are identical up to the $i^{th}$
state, at time $t_{k}$ there is an instance $I'$ of $A_{lazy'}$
with match and input buffers identical to those of $I$. Furthermore,
$s_{k}$ will also cause $I'$ to transition into $q_{i}^{lazy'}$.
From this point on, the processing of the two instances may diverge,
but we show that after the event $s_{l}$ is processed, both instances
will have transitioned into $q_{i+2}$ with the same input and match
buffers.

Any event corresponding to an event type already located in the match
buffer will be ignored by both $I$ and $I'$. Similarly, any event
corresponding to an event type requested by one of the later states
($q_{i+2}$ and beyond) will be stored to respective input buffers
of both instances. Thus, we are left to consider the events that trigger
the transitions $q_{i}^{lazy}\rightarrow q_{i+1}^{lazy}$ and $q_{i+1}^{lazy}\rightarrow q_{i+2}^{lazy}$.

We will examine the four possible scenarios for the above transitions
in $A_{lazy}$:
\begin{enumerate}
\item The transition $q_{i}^{lazy}\rightarrow q_{i+1}^{lazy}$ occurred
due to a corresponding event $s_{u}$ arriving from the input stream,
and the transition $q_{i+1}^{lazy}\rightarrow q_{i+2}^{lazy}$ occurred
due to a corresponding event $s_{v}$ taken from the input buffer.
Then we may conclude that $s_{v}$ arrived before $s_{u}$. Hence,
during evaluation in $A_{lazy'}$, after time $t_{k}$ the event $s_{v}$
is either already located in the input buffer, or will arrive eventually
before $s_{u}$, triggering the transition $q_{i}^{lazy'}\rightarrow q_{i+1}^{lazy'}$,
adding itself to the match buffer and removing itself from the input
buffer if taken from there. Then, the event $s_{v}$ will eventually
be received from the input stream and trigger the transition $q_{i+1}^{lazy'}\rightarrow q_{i+2}^{lazy'}$,
adding itself to the match buffer. Thus, at time $t_{l}$, both $A_{lazy}$
and $A_{lazy'}$ contain $s_{u}$ and $s_{v}$ in their respective
match buffers and do not contain $s_{v}$ in their input buffers.
\item The transition $q_{i}^{lazy}\rightarrow q_{i+1}^{lazy}$ occurred
due to a corresponding event $s_{u}$ taken from the input buffer,
and the transition $q_{i+1}^{lazy}\rightarrow q_{i+2}^{lazy}$ occurred
due to a corresponding event $s_{v}$ arriving from the input stream.
Then, during evaluation in $A_{sel'}$, the event $s_{u}$ is already
located in the input buffer by the time $s_{v}$ arrives from the
input stream. When $s_{v}$ arrives, it triggers the transition $q_{i}^{lazy'}\rightarrow q_{i+1}^{lazy'}$
and adds itself to the match buffer. Then, the event $s_{u}$ will
be immediately received from the input stream and will trigger the
transition $q_{i+1}^{lazy'}\rightarrow q_{i+2}^{lazy'}$, adding itself
to the match buffer and removing itself from the input buffer. Thus,
at time $t_{l}$, both $A_{lazy}$ and $A_{lazy'}$ contain $s_{u}$
and $s_{v}$ in their respective match buffers and do not contain
$s_{u}$ in their input buffers.
\item Both $q_{i}^{lazy}\rightarrow q_{i+1}^{lazy}$ and $q_{i+1}^{lazy}\rightarrow q_{i+2}^{lazy}$
were triggered by the corresponding events $s_{u},s_{v}$ arriving
on the input stream. Then we may conclude that $s_{u}$ arrived before
$s_{v}$. During evaluation in $A_{lazy'}$, the event $s_{u}$ will
be inserted to the input buffer upon its arrival, which will occur
after $t_{k}$. Later, the event $s_{v}$ will trigger the transition
$q_{i}^{lazy'}\rightarrow q_{i+1}^{lazy'}$, adding itself to the
match buffer. Immediately afterwards, the event $s_{u}$ located in
the input buffer will trigger $q_{i+1}^{lazy'}\rightarrow q_{i+2}^{lazy'}$,
adding itself to the match buffer and removing itself from the input
buffer. Thus, at time $t_{l}$, both $A_{lazy}$ and $A_{lazy'}$
contain $s_{u}$ and $s_{v}$ in their respective match buffers and
do not contain $s_{u}$ in their input buffers.
\item Both $q_{i}^{lazy}\rightarrow q_{i+1}^{lazy}$ and $q_{i+1}^{lazy}\rightarrow q_{i+2}^{lazy}$
were triggered by the corresponding events taken from the input buffer.
Then, during evaluation in $A_{lazy'}$, at time $t_{k}$ the events
which triggered both those transitions will already be located in
the input buffer (by the assumption above) and will trigger the transitions
$q_{i}^{lazy'}\rightarrow q_{i+1}^{lazy'}$ and $q_{i+1}^{lazy'}\rightarrow q_{i+2}^{lazy'}$
respectively, removing themselves from the input buffer and adding
themselves to the match buffer exactly as occurred during evaluation
in $A_{lazy}$. Thus, at time $t_{l}$, both $A_{lazy}$ and $A_{lazy'}$
contain $s_{u}$ and $s_{v}$ in their respective match buffers and
do not contain $s_{u}$ and $s_{v}$ in their input buffers.
\end{enumerate}
Thus, we have shown that in all cases the states of both $A_{lazy}$
and $A_{lazy'}$ are identical at $t_{l}$. As stated above, this
claim completes the proof.

\begin{lemma}

Let $P$ be a sequence pattern over event types $e_{1},\cdots,e_{k},\cdots,e_{n}$,
with a single negated event $e_{k}$. Let $A_{eager}$ be an eager
NFA and let $A_{lazy}$ be a Chain NFA derived from $P$. Then $A_{eager}$
and $A_{lazy}$ are equivalent.

\end{lemma}

We prove this lemma separately for two possible types of $A_{lazy}$,
Post-Processing Chain NFAs and First-Chance Chain NFAs.

Let us first consider the case in which $A_{lazy}$ is implemented
as a Post-Processing Chain NFA. Let $S=s_{1},...,s_{m}$ be a stream
of events. Assume without loss of generality that all events in $S$
are within the time window $W$. If $S$ does not contain an instance
of a negated event $e_{k}$, then, by Lemma 6.2, it is accepted by
$A_{eager}$ if and only if it is accepted by $A_{lazy}$. Otherwise,
we examine the two possible scenarios:
\begin{enumerate}
\item An event $s_{l}$ of type $e_{k}$ exists in $S$ and satisfies the
conditions required by the pattern. In this case, $A_{eager}$ will
discard all intermediate results and no match will be reported. In
$A_{lazy}$, by construction, a state $r$ exists, which is responsible
for detecting $e_{k}$. If the execution for all matches never reaches
$r$, then, since $r$ precedes $F$, no instance will reach the accepting
state and thus no match will be reported. Otherwise, if $r$ is reached
by some instance after $s_{l}$ has arrived, then $s_{l}$ will be
located in the input buffer, by construction of a Post-Processing
Chain NFA. Then, the edge $e_{r}^{take}$ will take $s_{l}$ and proceed
to the rejecting state, thus discarding the instance. Finally, if
$r$ is reached by some instance before $s_{l}$ has arrived, then
$e_{k}$ must appear at pattern end, after all positive events have
already arrived. By construction, \textit{r} will only have an outgoing
edge $e_{r}^{timeout}$ to the accepting state, and the resulting
\textit{search\_failed} event will not be processed. Following the
arrival of $s_{l}$, the edge $e_{r}^{take}$ will proceed to the
rejecting state, since conditions for $s_{l}$ are satisfied. Thus,
in any case, $A_{lazy}$ will discard all its partial matches.
\item An event $s_{l}$ of type $e_{k}$ exists in $S$ and does not satisfy
the conditions required by the pattern. In this case, $A_{eager}$
will ignore $s_{l}$ and report all matches. In $A_{lazy}$, by construction,
the edge $e_{r}^{take}$ of a state $r$ will never be traversed,
since its conditions do not hold for $s_{l}$. If $r$ is reached
by some instance after $s_{l}$ has arrived, then $s_{l}$ will be
located in the input buffer, and a search will trigger the \textit{search\_failed}
event, which will cause the instance to proceed to \textit{F}. Otherwise,
if $r$ is reached by some instance before $s_{l}$ has arrived, then
$e_{k}$ must appear at pattern end, after all positive events have
already arrived. By construction, \textit{r} will only have $e_{r}^{timeout}$
edge to the accepting state. When $s_{l}$ arrives, the edge $e_{r}^{take}$
will not be traversed, and the instance will only be affected by the
\textit{timeout} event, which will cause a transition to \textit{F}.
Thus, in any case, $A_{lazy}$ will report all its positive matches.
\end{enumerate}
To summarize, we have shown that in all cases $A_{eager}$ and $A_{lazy}$
report the same matches when applied on a stream \textit{S}.

Now, let $A_{lazy}$ be implemented as a First-Chance Chain NFA. Again,
let $S=s_{1},...,s_{m}$ be a stream of events and assume w.l.o.g.
that all events in $S$ are within the time window $W$. If $S$ does
not contain an instance of a negated event $e_{k}$, then, by Lemma
6.2, it is accepted by $A_{eager}$ if and only if it is accepted
by $A_{lazy}$. Otherwise, we examine two possible scenarios:
\begin{enumerate}
\item An event $s_{l}$ of type $e_{k}$ exists in $S$ and satisfies the
conditions required by the pattern. In this case, $A_{eager}$ will
discard all intermediate results and no match will be reported. In
$A_{lazy}$, by construction, a \textit{take} edge $e_{k}^{take}$
exists, detecting an event of type $e_{k}$ and proceeding to the
rejecting state \textit{R}. Let \textit{q} denote the source state
of $e_{k}^{take}$. If the execution for all matches never reaches
\textit{q}, then, since \textit{q} precedes $F$, no instance will
reach the accepting state and no match will be reported. Otherwise,
if \textit{q} is reached by some instance after $s_{l}$ has arrived,
then $s_{l}$ will be located in the input buffer, by construction
of $A_{lazy}$. Then, $e_{k}^{take}$ will take $s_{l}$ and proceed
to \textit{R}. Note that \textit{q} cannot be reached before $s_{l}$
has arrived, since in this case $e_{k}$ must appear at the pattern's
end (by construction, requiring the positive event succeeding $e_{k}$
to be taken prior to entering \textit{q}), and this type of pattern
is not supported by First-Chance Chain NFAs. Thus, in any case, $A_{lazy}$
will discard all its partial matches.
\item An event $s_{l}$ of type $e_{k}$ exists in $S$ and does not satisfy
the conditions required by the pattern. In this case, $A_{eager}$
will ignore $s_{l}$ and report all matches. In $A_{lazy}$, by construction,
$e_{k}^{take}$ will never be traversed, since its conditions do not
hold for $s_{l}$. With the exception of $e_{k}^{take}$, the structure
of $A_{lazy}$ is identical to that of a lazy sequence Chain NFA accepting
a pattern P with $e_{k}$ ignored. Therefore, by Lemma 6.2, all matches
reported by $A_{eager}$ will also be reported by $A_{lazy}$.
\end{enumerate}
Thus, we have shown that in all cases $A_{eager}$ and $A_{lazy}$
report the same matches when applied on a stream \textit{S}, which
completes the proof.

\newtheorem{corollary}{Corollary}
\begin{corollary}

Let $P$ be a sequence pattern over event types $e_{1},\cdots,e_{n}$,
with an arbitrary number $1\leq m<n$ of negated events. Let $A_{eager}$
be an eager NFA and let $A_{lazy}$ be a Chain NFA derived from $P$.
Then $A_{eager}$ and $A_{lazy}$ are equivalent.

\end{corollary}

The proof is by induction on $m$, with Lemma 6.3 serving as an induction
basis. The induction step for $m=i+1$ is proven in the same way as
Lemma 6.3, using the induction hypothesis instead of Lemma 6.2 for
assuming the correctness of the pattern without the $m^{th}$ negated
event.

\begin{lemma}

Let $P$ be a sequence pattern over event types $e_{1},\cdots,e_{k},\cdots,e_{n}$,
with a single iterated event $e_{k}$. Let $A_{eager}$ be an eager
NFA and let $A_{lazy}$ be a Chain NFA derived from $P$. Then $A_{eager}$
and $A_{lazy}$ are equivalent.

\end{lemma}

We will prove this lemma by double inclusion, i.e., by showing that,
given a potential match $M=s_{1},...,s_{m}$ for pattern \textit{P},
$A_{eager}$ accepts \textit{M} if and only if $A_{lazy}$ does. This,
in turn, will be proven by induction on the number of events of type
$e_{k}$ in \textit{M}.

For the induction basis, assume that \textit{M} contains only a single
event of type $e_{k}$. Then, let \textit{P'} be a pattern identical
to \textit{P}, but with $e_{k}$ as a non-iterated type. By Lemma
6.2, if \textit{M} is a match for \textit{P}, then it is also a match
for \textit{P'}. By iteration definition, both $A_{eager}$ and $A_{lazy}$
also accept \textit{P'}. Hence, either both NFA accept the match \textit{M},
or both reject it.

For the induction step, assume that the condition holds for up to
\textit{i} events of type $e_{k}$, and let \textit{M} contain \textit{i+1}
events of type $e_{k}$: $s_{1}^{k},...,s_{i}^{k},s_{i+1}^{k}$.

Let \textit{M} be accepted by $A_{eager}$. Let $t_{i+1}$ be the
arrival time of $s_{i+1}^{k}$. Then, a state \textit{q} exists in
$Q_{eager}$, such that at time $t_{i+1}$ the NFA instance of $A_{eager}$
is in $q$. Since $i\geq1$, \textit{q} is a state containing a self-loop
taking $e_{k}$. Hence, before and after $t_{i+1}$, the current state
was q. Since \textit{M} is accepted, we know that all events that
arrived after $t_{i+1}$ caused the NFA instance to reach \textit{F}
from \textit{q}. Let \textit{M'} be a match identical to \textit{M},
but without $s_{i+1}^{k}$. By the observation above, \textit{M'}
is also accepted by $A_{eager}$. By the induction hypothesis, \textit{M'}
is thus accepted by $A_{lazy}$. Now, since \textit{M} is accepted
by $A_{eager}$, it follows that $s_{i+1}^{k}$ satisfies the conditions
with other events in \textit{M}. Therefore, when an \textit{iterate}
edge in $A_{lazy}$ fetches a subset $s_{1}^{k},...,s_{i+1}^{k}$,
it is added to the match buffer. By construction of a Lazy Iteration
Chain NFA, the edge is traversed, reaching \textit{F}. Hence, \textit{M}
is accepted by $A_{lazy}$.

Now, let M be accepted by $A_{lazy}$. Consequently, when a subset
$s_{1}^{k},...,s_{i+1}^{k}$ is fetched by an iterate edge of $A_{lazy}$,
all events in this subset satisfy the conditions with other events.
In particular, it follows that a subset $s_{1}^{k},...,s_{i}^{k}$
satisfies the conditions as well. It also follows that during evaluation
of \textit{M} some NFA instance reaches the state with an outgoing
\textit{iterate} edge. Hence, during the same traversal which retrieves
$s_{1}^{k},...,s_{i+1}^{k}$, a subset $s_{1}^{k},...,s_{i}^{k}$
is also generated and causes the transition to occur. That is, the
match \textit{M'} containing $s_{1}^{k},...,s_{i}^{k}$ without $s_{i+1}^{k}$
is also accepted by $A_{lazy}$. By the induction hypothesis, \textit{M'}
is thus accepted by $A_{eager}$. By definition of the eager NFA,
after $s_{i}^{k}$ arrives during evaluation, an instance exists whose
current state is a state \textit{q} with a self-loop $e_{loop}$ taking
$e_{k}$. When $s_{i+1}^{k}$ arrives, this instance will thus attempt
a traversal of $e_{loop}$. By assumption, $s_{1}^{k},...,s_{i+1}^{k}$
satisfies all the conditions required by the pattern, hence this transition
will succeed. Since \textit{M'} is accepted, we know that all events
that arrived after $s_{i}^{k}$ in \textit{M'} caused the NFA instance
to reach \textit{F} from \textit{q}. Consequently, the same event
sequence will cause the same transitions in \textit{M}, i.e., \textit{M}
is accepted by $A_{eager}$.

To summarize, we have shown that, given a sequence pattern \textit{M}
containing an iterated event, $A_{eager}$ accepts \textit{M} if $A_{lazy}$
does, and vice versa, which proves their equivalence.

\begin{corollary}

Let $P$ be a sequence pattern over event types $e_{1},\cdots,e_{n}$,
with an arbitrary number $1\leq m<n$ of iterated events. Let $A_{eager}$
be an eager NFA and let $A_{lazy}$ be a Chain NFA derived from $P$.
Then $A_{eager}$ and $A_{lazy}$ are equivalent.

\end{corollary}

The proof is identical to that of Corollary 6.4, using Lemma 6.5 instead
of Lemma 6.3.

\newtheorem{definition}{Definition}
\begin{definition}

Let $A=\left(Q,E,q_{1},F,R\right)$ be an NFA. We will call \textit{A}
a \textbf{composite NFA of size }\textbf{\textit{K}}, if there exist
automata $A_{1},\cdots A_{k}$, such that \textit{A} consists of the
union of them, with their respective initial states merged into $q_{1}$,
the accepting states merged into $F,$ and the rejecting states merged
into $R$.

\end{definition}

\begin{lemma}

Let $A_{1}$,$A_{2}$ be two composite NFAs of size \textit{K}, i.e.,
$A_{1}$ consists of $A_{1}^{1},\cdots A_{k}^{1}$ and $A_{2}$ consists
of $A_{1}^{2},\cdots A_{k}^{2}$. Then, if the sub-automata of $A_{1}$
and $A_{2}$ are pairwise equivalent, i.e., $\left(A_{1}^{1}\equiv A_{1}^{2}\right)\wedge\cdots\wedge\left(A_{k}^{1}\equiv A_{k}^{2}\right)$,
then $A_{1}$ and $A_{2}$ are equivalent.

\end{lemma}

The proof is by induction on \textit{K}. For $K=1$, the correctness
follows immediately, since $A_{1}=A_{1}^{1}\equiv A_{1}^{2}=A_{2}$.
For the induction step, assume the claim to hold for $K=i$ and let
$A_{1}$,$A_{2}$ be two composite NFAs of size $i+1$. Now, assume
that all sub-automata of $A_{1}$ and $A_{2}$ are equivalent. We
will show the equivalence of $A_{1}$ and $A_{2}$ by mutual inclusion.

Let $M=s_{1},...,s_{m}$ be a potential match for $A_{1}$,$A_{2}$.
Examine the following possible situations:
\begin{enumerate}
\item \textit{M} is accepted by $A_{i+1}^{1}$. Then, by construction of
$A_{1}$, it also accepts \textit{M}. On the other side, by assumption
of equivalence of $A_{i+1}^{1}$ and $A_{i+1}^{2}$, \textit{M} is
also accepted by $A_{i+1}^{2}$. By construction of $A_{2}$, it also
accepts \textit{M}.
\item \textit{M} is accepted by $A_{1}$, but not accepted by $A_{i+1}^{1}$.
Let $\bar{A_{1}}$,$\bar{A_{2}}$ be constructed by removing all the
internal states of $A_{i+1}^{2}$ and $A_{i+1}^{2}$, respectively.
\textit{M} is accepted by $\bar{A_{1}}$, according to the assumption.
Then, by the induction hypothesis, \textit{M} is also accepted by
$\bar{A_{2}}$, since $\bar{A_{1}}$,$\bar{A_{2}}$ are composite
NFAs of size \textit{i}. By construction of $\bar{A_{2}}$, M is also
accepted by $A_{2}$.
\item \textit{M} is not accepted by $A_{1}$. Then, in particular, it is
not accepted by any of its sub-automata $A_{1}^{1},\cdots A_{i}^{1},A_{i+1}^{1}$.
By assumption of equivalence, \textit{M} is also not accepted by any
of $A_{1}^{2},\cdots A_{i}^{2},A_{i+1}^{2}$. By construction of $A_{2}$,
it also does not accept \textit{M}.
\end{enumerate}
To summarize, we have shown that, for any potential match \textit{M},
it is either accepted by both $A_{1}$ and $A_{2}$ or by neither
of them, which completes the proof.

\begin{lemma}

Let $P$ be a partial sequence pattern over event types $e_{1},\cdots,e_{n}$,
with possible negated or iterated events. Let $A_{chain}$ be a Chain
NFA for \textit{P}, as presented in Section \ref{sub:Partial-Sequence}.
Let $SEQ=\left\{ seq_{1},\cdots,seq_{k}\right\} $ be a set of all
sequence patterns whose orders are allowed by \textit{P} (i.e., \textit{P}
is a union of patterns in \textit{SEQ}). Finally, let $A_{multi-chain}$
be a Lazy Multi-Chain NFA (Section \ref{sub:Disjunction}), with each
sub-chain being a Chain NFA for a sequence pattern in \textit{SEQ}
(Section \ref{sub:Sequence}). Then $A_{chain}$ and $A_{multi-chain}$
are equivalent.

\end{lemma}

The proof is by mutual inclusion. Let $M=s_{1},...,s_{m}$ be a potential
match for $A_{1}$,$A_{2}$. Examine the following scenarios:
\begin{enumerate}
\item \textit{M} is accepted by $A_{multi-chain}$. Then, in particular,
it is accepted by at least one of its sub-chains. By construction,
this Chain NFA $A_{i}$ and $A_{chain}$ are identical except for
$A_{i}$ having stricter ordering constraints. Consequently, \textit{M}
is also accepted by $A_{chain}$.
\item \textit{M} is rejected by $A_{multi-chain}$. Then, it is also rejected
by each of its sub-chains. Observe that the only difference between
$A_{chain}$ and each of the sub-chains is the ordering constraint
definition. Hence, the reason for the rejection can only be related
to violating temporal conditions. Assume w.l.o.g. that an event $s_{p}$
of type $e_{u}$ in \textit{M} precedes an event $s_{r}$ of type
$e_{v}$, which is forbidden by all chains. Also, assume that $e_{v}$
precedes $e_{u}$ in the frequency order (the opposite case is symmetrical).
Then, as $A_{chain}$ reaches a state responsible for taking $e_{u}$,
$s_{r}$ is already located in the match buffer. By definition of
an ordering filter $succ_{i}$ for Chain NFA for partial sequences,
$succ_{i}$ will contain $e_{v}$, hence $s_{p}$ will not be taken
by $A_{chain}$, discarding \textit{M}.
\end{enumerate}
\begin{corollary}

Let $P$ be a conjunction pattern over event types $e_{1},\cdots,e_{n}$,
with possible negated or iterated events. Let $A_{chain}$ be a Chain
NFA for \textit{P}. Let us define $SEQ=\left\{ seq_{1},\cdots,seq_{k}\right\} $
as a set of all sequences over $e_{1},\cdots,e_{n}$. Let $A_{multi-chain}$
be a Lazy Multi-Chain NFA with each sub-chain being a Chain NFA for
a sequence pattern in \textit{SEQ}. Then $A_{chain}$ and $A_{multi-chain}$
are equivalent.

\end{corollary}

The proof immediately follows from Lemma 6.9, since a conjunction
pattern is an edge case of a partial sequence, with no ordering constraints
between the primitive events.

\begin{corollary}

Let $P$ be a partial sequence pattern over $e_{1},\cdots,e_{n}$.
Let $A_{eager}$ be an eager NFA and let $A_{lazy}$ be a Chain NFA
derived from $P$. Then $A_{eager}$ and $A_{lazy}$ are equivalent.

\end{corollary}

Assume w.l.o.g. that $A_{eager}$ is implemented as a composite NFA,
containing a sub-automaton for each valid sequence in \textit{P}.
An NFA with this structure necessarily exists, since each partial
sequence can be represented as a disjunction of sequences. Let $A_{multi-chain}$
be a Lazy Multi-Chain NFA with each sub-chain being a Chain NFA for
each valid sequence in \textit{P}. By Lemma 6.8, $A_{eager}$ and
$A_{multi-chain}$ are equivalent. By Lemma 6.9, $A_{multi-chain}$
and $A_{lazy}$ are equivalent. Then, by transitivity, $A_{eager}$
and $A_{lazy}$ are equivalent.

\begin{corollary}

Let $P$ be a conjunction pattern over event types $e_{1},\cdots,e_{n}$.
Let $A_{eager}$ be an eager NFA and let $A_{lazy}$ be a Chain NFA
derived from $P$. Then $A_{eager}$ and $A_{lazy}$ are equivalent.

\end{corollary}

The proof immediately follows from Corollary 6.11.

\begin{lemma}

Let $P$ be a disjunction pattern over $e_{1},\cdots,e_{n}$, i.e.,
$P=OR\left(p_{1},\cdots p_{m}\right)$, where each $p_{i}$ is a pattern
over sequence, conjunction, partial sequence, negation and iteration
operators. Let $A_{eager}$ be an eager NFA and let $A_{lazy}$ be
a Lazy Multi-Chain NFA derived from $P$. Then $A_{eager}$ and $A_{lazy}$
are equivalent.

\end{lemma}

Assume w.l.o.g. that $A_{eager}$ is implemented as a composite NFA,
containing a sub-automaton for each sub-pattern $p_{i}$. Then, $A_{eager}$
and $A_{lazy}$ are both composite NFA of size \textit{m}. By definition
of $p_{i}$, by Lemmas 6.2, 6.3 and 6.5 and by Corollaries 6.11 and
6.12, each sub-automaton for $p_{i}$ in $A_{eager}$ is equivalent
to a corresponding sub-chain in $A_{lazy}$. Then, by Lemma 6.8, $A_{eager}$
and $A_{lazy}$ are equivalent.

We will now complete the proof of Theorem 6.1. If \textit{P} does
not contain the disjunction operator, the equivalence of $A_{eager}$
and $A_{lazy}$ follows from Lemmas 6.2, 6.3 and 6.5 and from Corollaries
6.11 and 6.12. Otherwise, let \textit{P'} be the DNF form of pattern
\textit{P}. It is sufficient to prove that the claim holds for \textit{P'}.
By definition of DNF, \textit{P'} is a disjunction pattern. Thus,
by Lemma 6.13, the corresponding eager and lazy NFA for \textit{P'}
are equivalent.$\blacksquare$

\section{Lazy Tree NFA}

\label{sec:Lazy-Tree-NFA}

The Chain NFA described in Section \ref{sec:Lazy-Chain-NFA} may significantly
improve evaluation performance, provided we know the correct frequency
order. As shown in the examples above, the more drastic the difference
between the arrival rates of different events, the greater the potential
improvement.

There are, however, several drawbacks which severely limit the applicability
of the Chain NFA in real-life scenarios. First, the assumption of
specifying the frequency order in advance is not always realistic.
In many cases, it is hard or even impossible to predict the actual
arrival rates of primitive event types. Note that the described model
is very sensitive to wrong guesses, as specifying a low-frequency
event before a high-frequency event will yield many redundant evaluations
and overall poor performance. Second, even if it is possible to set
up the system with a correct frequency order, we can rarely guarantee
that it will remain the same during the run. In many real-life applications
the data is highly dynamic, and arrival rates of different events
are subject to change on-the-fly. Such diversity may cause an initially
efficient Chain NFA to start performing poorly at some point. Continual
changes may come, for example, in the form of bursts of usually rare
events.

To overcome these problems we introduce the notion of ad hoc order.
Instead of relying on a single frequency order specified at the beginning
of the run, we determine the current arrival rates on-the-fly and
modify the actual evaluation chain according to the order reflecting
the current frequencies of the events. Our NFA will thus have a tree
structure, with each of its nodes (states) ``routing'' the incoming
events to the next ``hop'' according to this dynamically changing
order. By performing these ``routing decisions'' at each evaluation
step, we guarantee that any partial match will be evaluated using
the most efficient order possible at the moment.

To implement the desired functionality, we require that each state
have knowledge regarding the current frequency of each event type.
We will use the input buffer introduced above to this end. By its
definition, the input buffer of a particular NFA instance contains
all events that arrived from the input stream within the specified
time window. For each event type, we will introduce a counter containing
the current number of events matched with this event type inside the
buffer. This counter will be incremented on each insertion of a new
event with the corresponding type and decremented upon its removal. 

Matching the pattern requires at least one event corresponding to
each event type to be present in the input buffer. Hence, we will
add a condition stating that no evaluation will be made by a given
instance until all the counters are greater than zero. Only when all
of the event counters are greater than zero does it make sense to
determine the evaluation order, since otherwise the missing event(s)
may not arrive at all and the partial matching process will be redundant.
After the above condition is satisfied, we can derive the exact ascending
order of frequencies based on the currently available data by sorting
the counters.

The above calculation will be performed by each state on each matching
attempt, and the resulting value will be used to determine the next
step in the evaluation order. In terms of NFA, this means that a state
needs to select the next state for a partial match based on the current
contents of the input buffer. To this end, a state has several outgoing
\textit{take} edges as opposed to a single one in Chain NFA. Each
edge takes a different event type and the edges point to different
states. We will call the NFA employing this structure a \textit{Tree
NFA} and will formally define this model below.

We will illustrate the above using the following example. Consider
the stock trading scenario from Example 1.1. Assume now that all stock
updates are represented using a single event type, which we will denote
by \textit{E}. This event type has two data attributes: a categorical
attribute called ``ticker,'' which represents the stock for which
the event has occurred, and a numerical attribute called ``price,''
which is the price of the stock. The arrival rates of events with
a particular values for ``ticker'' attribute (i.e., events corresponding
to updates of a specific stock) are unknown and subject to change.
We are interested in monitoring the following SASE+ pattern:

\begin{equation} 
\begin{array}{l}
PATTERN\: SEQ\left(E\: a,E\: b,E\: c\right)\\ 
WHERE\: skip\_till\_any\_match\:\{\\ 
\qquad a.ticker = MSFT\\ 
\qquad AND\: b.ticker = GOOG\\ 
\qquad AND\: c.ticker = AAPL\\ 
\qquad AND\: a.price < b.price\\ 
\qquad AND\: b.price < c.price\}\\ 
WITHIN\:1\: hour. 
\end{array} 
\end{equation} 

Here, MSFT, GOOG and AAPL stand for the stocks of Microsoft, Google
and Apple respectively.

\begin{figure}   \centering \includegraphics[width=\linewidth]{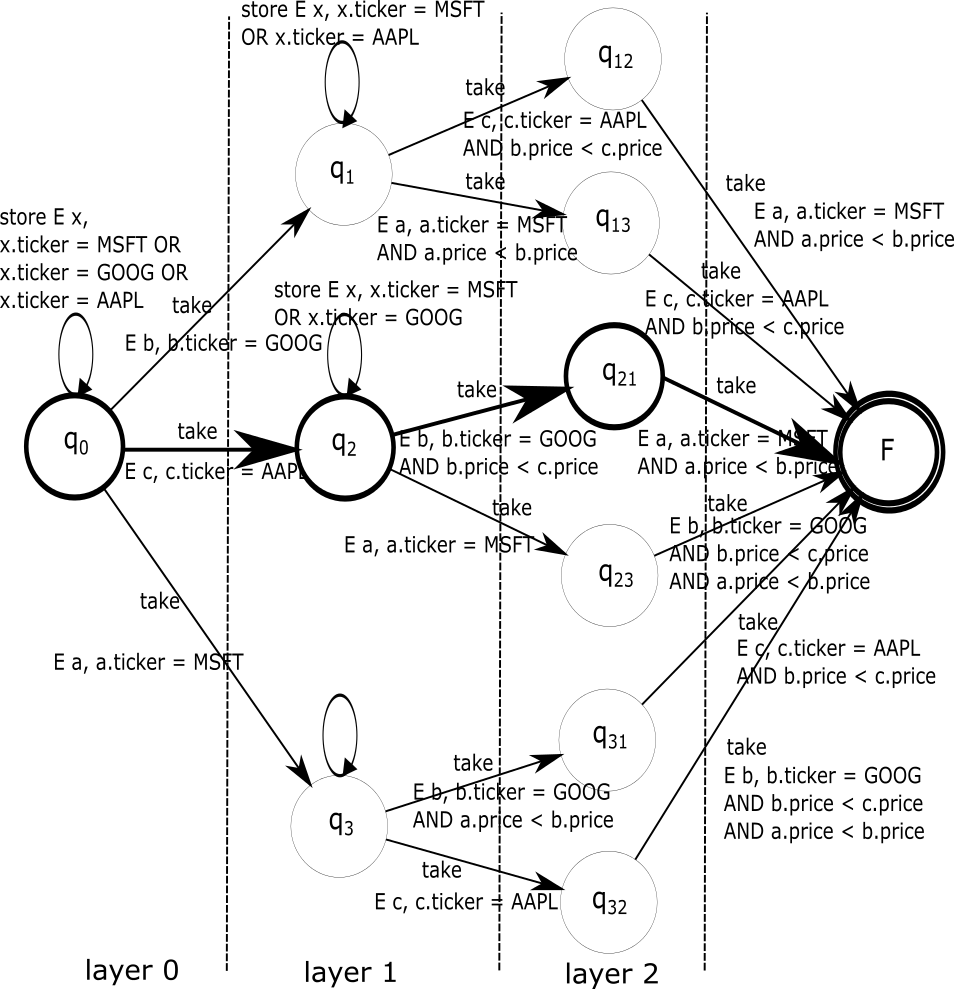}   \caption{Tree NFA example.} \label{fig:Tree-NFA} \end{figure}

Figure \ref{fig:Tree-NFA} illustrates a Tree NFA for the this pattern.
For simplicity, ignore and timeout edges are omitted.

In formal terms, a Tree NFA is structured as a tree of depth $n-1$,
the root being the initial state and the leaves connected to the accepting
state. Nodes located at each layer $k;\:0\leq k\leq n-1$ (i.e., all
nodes in depth $k$) are all states responsible for all orderings
of $k$ event types out of the $n$ event types defined in the pattern.
Each such node has $n-k$ outgoing edges, one for each event type
which does not yet appear in the partial ordering this node is responsible
for. Those edges are connected to states at the next layer, responsible
for all extensions of the ordering of this particular node to length
of $k+1$. The only exceptions to this rule are the leaves, which
have a single outgoing edge, connected directly to the final state.

For instance, in the example in Figure \ref{fig:Tree-NFA}, layer
0 contains the initial state $q_{0}$, layer 1 contains states $q_{1},q_{2},q_{3}$,
and layer 2 contains the states $q_{12},q_{13},q_{21},q_{23},q_{31},q_{32}$.

More formally, the states for a Tree NFA are defined as follows. Let
$O_{k}$ denote the ordered subsets of size $k$ of the event types
$e_{1},\cdots,e_{n}$. Let
\[
Q_{k}=\left\{ q_{ord}|ord\in O_{k}\right\} 
\]
denote the set of states at the layer $k$ (note that $Q_{0}=\left\{ q_{0}\right\} $).
Then the set of all states of the Tree NFA is
\[
Q=\bigcup_{k=0}^{n-1}Q_{k}\cup\left\{ F,R\right\} 
\]
\[
q_{0}=q_{\oslash}.
\]

To describe the edges and their respective conditions, some preliminary
definitions are needed.

First, we will complete the definitions required for the ordering
filters. Since each state $q_{ord}$ corresponds to some evaluation
order prefix $ord$, we will set $ord_{e}=ord$ for each outgoing
edge $e$ of $q_{ord}$. As mentioned earlier, it is enough for $ord_{e}$
to be a partial order ending with $\hat{e}$. In other words, each
$take$ edge in the tree derives the corresponding filter for its
target event type from the order used for reaching this edge.

Let $c_{e}$ denote the value of the counter of events associated
with the type $e$ in the input buffer. Let $se\left(q_{ord}\right)=min\left(\left\{ c_{e}|e\notin ord\right\} \right)$
denote the most infrequent event in the input buffer during the evaluation
step in which $q_{ord}$ is the current state. Finally, we will define
the predicate $p_{ne}\left(q_{ord}\right)$ (non-empty) as the condition
on the input buffer of state $q_{ord}$ to contain at least a single
instance of each primitive event not appearing in $ord$ and another
predicate $p_{se}\left(q_{ord},e\right)$ to be true if and only if
an event $e$ corresponds to event type $se\left(q_{ord}\right)$.
Let $E_{ord}$ denote the set of outgoing edges of $q_{ord}$. Then,
$E_{ord}$ will contain the following edges:
\begin{itemize}
\item $e_{ord}^{ignore}=\left(q_{ord},q_{ord},ignore,ord,true,\varnothing,\varnothing\right)$:
any event whose type corresponds to one of the already taken events
(appearing in the ordering this state corresponds to) is ignored.
\item For each primitive event $e\notin ord$:

\begin{itemize}
\item $e_{ord,e}^{store}=\left(q_{ord},q_{ord},store,e,\neg p_{ne}\left(q_{ord}\right)\vee\neg p_{se}\left(q_{ord},e\right),\varnothing,\varnothing\right)$:
when either the $p_{ne}$ or $p_{se}$ condition is not satisfied,
the incoming event is stored into the input buffer.
\item %
{} \resizebox{\linewidth}{!} {$e_{ord,e}^{take}=\left(q_{ord},q_{ord,e},take,e,p_{ne}\left(q_{ord}\right)\wedge p_{se}\left(q_{ord},e\right)\wedge cond_{e},prec_{e},succ_{e}\right)$}:
if the contents of the input buffer satisfy the $p_{ne}$ and $p_{se}$
predicates and an incoming event with a type $e$ (1) satisfies the
conditions required by the initial pattern (denoted by $cond_{e}$);
and (2) is located within the scope defined for this state, it is
taken into the match buffer and the NFA instance advances to the next
layer of the tree.
\end{itemize}
\item For states in the last layer (where $|ord|=n$), the $take$ edges
are of the form 
\[
e_{ord,e}^{store}=\left(q_{ord},F,take,e,p_{ne}\left(q_{ord}\right)\wedge cond_{e}\wedge InScope_{ord}\left(e\right),prec_{e},succ_{e}\right).
\]

\end{itemize}
The ordering filters $\left(prec_{e},succ_{e}\right)$ for a Tree
NFA are calculated the same way as for a Chain NFA, as described in
Section \ref{sec:Lazy-Chain-NFA}. However, since the Tree NFA does
not have a predefined frequency order \textit{freq} to be used for
calculating the ordering filters, $ord_{e}$ will be used instead.
This order is the effective frequency order applied on the current
input.

In addition, each state features a $timeout$ edge similarly to the
one described for the Chain NFA. We will denote the set of all timeout
edges as $E_{timeout}$.

The set of all edges for Tree NFA is defined as follows:
\[
E=\left(\bigcup_{\left\{ ord|q_{ord}\in Q\right\} }E_{i}\right)\cup E_{timeout},
\]

and the NFA itself is defined as follows:
\[
A=\left(Q,E,q_{1},F,R\right),
\]
where $Q$ and $E$ are as defined above.

It can be observed that a Tree NFA contains all the possible Chain
NFAs for a given pattern, with shared states for common subpatterns.
Thus, the execution of a Tree NFA on any input is equivalent to the
execution of some Chain NFA on that input. The conditions on Tree
NFA edges are designed in such a way that the least frequent event
is chosen at each evaluation step. Hence, this Chain NFA is always
the one whose given frequency order is the actual frequency order
as observed from the input stream. An example is shown in Figure \ref{fig:Tree-NFA}.
Nodes and edges marked in bold illustrate the evaluation path for
an input stream satisfying $count(AAPL)\leq count(GOOG)\leq count(MSFT)$,
i.e., corresponding to the frequency order \textit{c,b,a}.

When implementing the Tree NFA, the number of states might be exponential
in $n$. To overcome this limitation, we propose to implement lazy
instantiation of NFA states \textendash{} only those states reached
by at least a single active instance will be instantiated and will
actually occupy memory space. After all NFA instances reaching a particular
state are terminated, the state will be removed from the NFA as well.
Even though the worst case complexity remains exponential in this
case, in practice there will be fewer changes in the event rates than
there will be new instances created. This conclusion is supported
by our experiments, which are explained in the next section.

For disjunctions and composite patterns, the solution described above
can be extended to a \textit{Multi-Tree NFA} similarly to the way
a Chain NFA was extended to a Multi-Chain NFA in Section \ref{sec:Lazy-Multi-Chain-NFA}.
In this topology, the initial state contains an outgoing $take$ edge
for each of the disjunction clauses in a pattern, leading to the dedicated
Tree NFA corresponding to this clause.

\section{Experimental Evaluation}

\label{sec:Experimental-Evaluation}

This section presents a detailed experimental analysis of our method.
We applied Lazy (Multi-)Chain and Tree NFA on wide range of patterns,
assessing their efficiency and scalability. A comparison with an eager
NFA-based CEP framework was conducted, demonstrating an overall performance
gain of up to two orders of magnitude in most scenarios.

Our metrics for this study are throughput, memory consumption, and
runtime complexity. Throughput is defined as the number of events
processed per second. For memory consumption, we estimate the peak
size (in MB) of the reserved system memory. Runtime complexity is
measured as the number of calculations and memory modifications per
event. For convenience, these two types of operations were evaluated
separately.

Real-world historical stock price data from the NASDAQ stock market,
taken from \cite{EODData}, was used in the experiments. This data
spans a 1-year period, covering over 2100 stock identifiers with prices
updated on a per minute basis. Our input stream contained 80,509,033
primitive events, each consisting of a stock identifier, a timestamp,
and a current price. We augmented the event format with additional
attributes, namely a history list of older prices and an identifier
of a region to which the stock belongs, e.g., North America or Europe.
For each of the 8 regions defined by NASDAQ, a corresponding event
type was defined. A primitive event was then considered to belong
to a type corresponding to the region of its stock identifier.

The patterns used during evaluation differed only in their operator
(e.g., SEQ or AND), the participating event types (i.e., regions)
and the time window size. All patterns shared an identical condition
of high correlation between each pair of primitive events. That is,
the Pearson correlation coefficient between price history lists was
required to exceed a predefined threshold \textit{T}. For example,
for a sequence operator, three event types, \textit{NACompany}, \textit{EuCompany},
\textit{AfrCompany}, and a time window \textit{h}, the corresponding
pattern would be declared as follows:

\begin{equation} 
\begin{array}{l} 
PATTERN\: SEQ\\ 
\qquad \left(NACompany\: a,EUCompany\: b,AfrCompany\: c\right)\\ 
WHERE\: skip\_till\_any\_match\:\{\\ 
\qquad Corr\left(a.history,b.history\right)>T\\ 
\qquad AND\: Corr\left(b.history,c.history\right)>T\\ 
\qquad AND\: Corr\left(c.history,a.history\right)>T\}\\ 
WITHIN\: h. 
\end{array} 
\end{equation}

The following pattern types were used during this study:
\begin{itemize}
\item Sequences of 3, 4 and 5 primitive events (marked on all graphs as
SEQ3, SEQ4 and SEQ5, respectively).
\item Conjunctions of 2 and 3 primitive events (marked as AND2, AND3). Due
to the extremely rapid growth of NFA instances during eager evaluation
attempts, experiments for higher numbers of events could not be conducted
in the environment available for this research.
\item Negation patterns for SEQ3, AND3 and SEQ5. In all of the above, the
second event was negated. For SEQ5 alone, we evaluated both methods
for lazy negation presented in Section \ref{sub:Negation}. They are
marked as Lazy-PP and Lazy-FC on all graphs. For the rest of the patterns,
only Lazy Post-Processing Chain NFA was used.
\item Iteration pattern of 3 events (i.e., SEQ(a,b+,c)), with the second
event being iterated (marked as ITER3). Due to the exponential number
of instances and limited hardware resources, experiments for this
pattern were conducted on time windows half the size of those used
for other pattern types.
\item Disjunction patterns consisting of: (1)two sequences of two events
(OR2SEQ2); (2)four sequences of two events (OR4SEQ2); (3)two sequences
of four events (OR2SEQ4).
\end{itemize}
All NFA models under examination were implemented in Java. The experiments
were run on a machine with 2.20 Ghz CPU and 16.0 GB RAM.

\subsection{Time Window Size}

\label{sub:Time-Window-Size}

In our first set of experiments, we compared how the lazy and the
eager strategies scale with a growing time window size. Both algorithms
were repeatedly applied on a data stream, with values for \textit{h}
ranging between 5 and 30 minutes. For the iteration pattern alone,
the range was modified to 2-12 minutes.

\begin{figure*}   \centering \includegraphics[width=.8\linewidth]{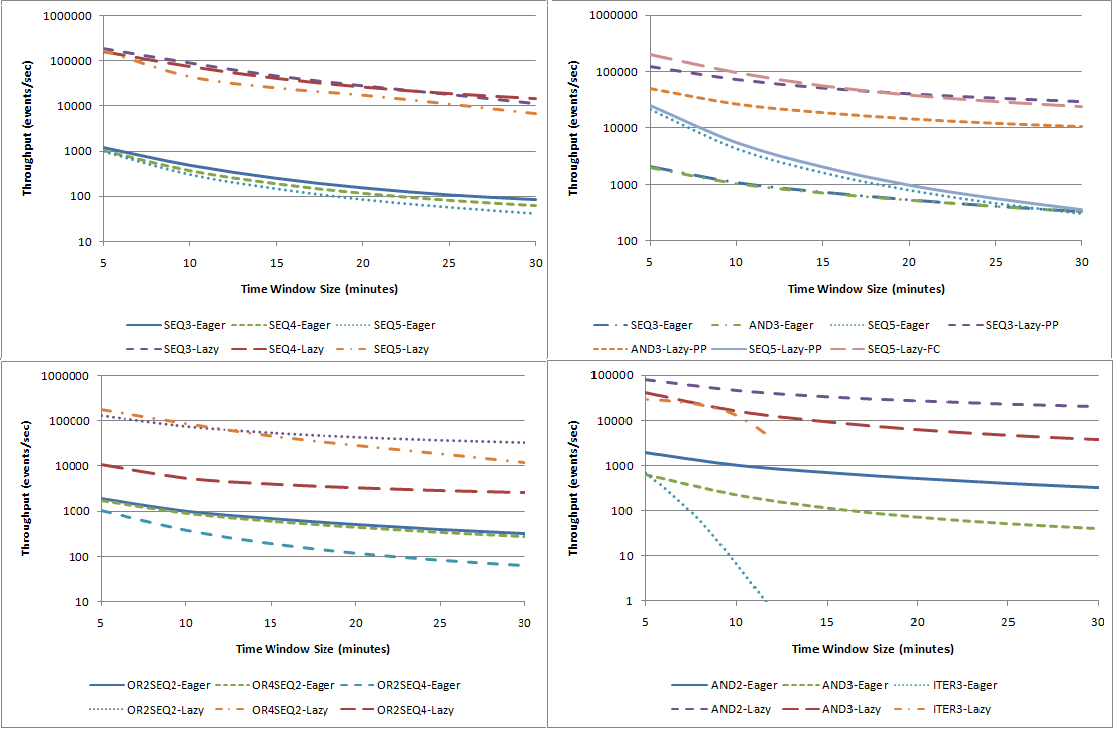}   \caption{Throughput as a function of time window size for various patterns (logarithmic scale).} \label{fig:Throughput} \end{figure*}

All figures summarizing this experiment contain several sub-graphs,
presenting results for different pattern types as listed above.

Figure \ref{fig:Throughput} describes the throughput as a function
of \textit{h}. For presentation clarity, the logarithmic scale was
used. A steady increase of one to two orders of magnitude can be observed
for the lazy approach as compared to its eager counterpart. The performance
decline for larger values of \textit{h} is generally smoother for
lazy NFA or comparable in both models. For the negated sequence of
five events, a clear advantage of the First-Chance Chain NFA over
the Post-Processing method can be seen. This holds due to the negated
event being the second in the sequence. Hence, according to First-Chance
strategy, verification of its absence can occur as the third event
is accepted, rather than after all positive events have arrived. For
disjunctions, sub-sequence length affects the overall throughput significantly
more than the number of sub-sequences.

\begin{figure*}   \centering \includegraphics[width=.8\linewidth]{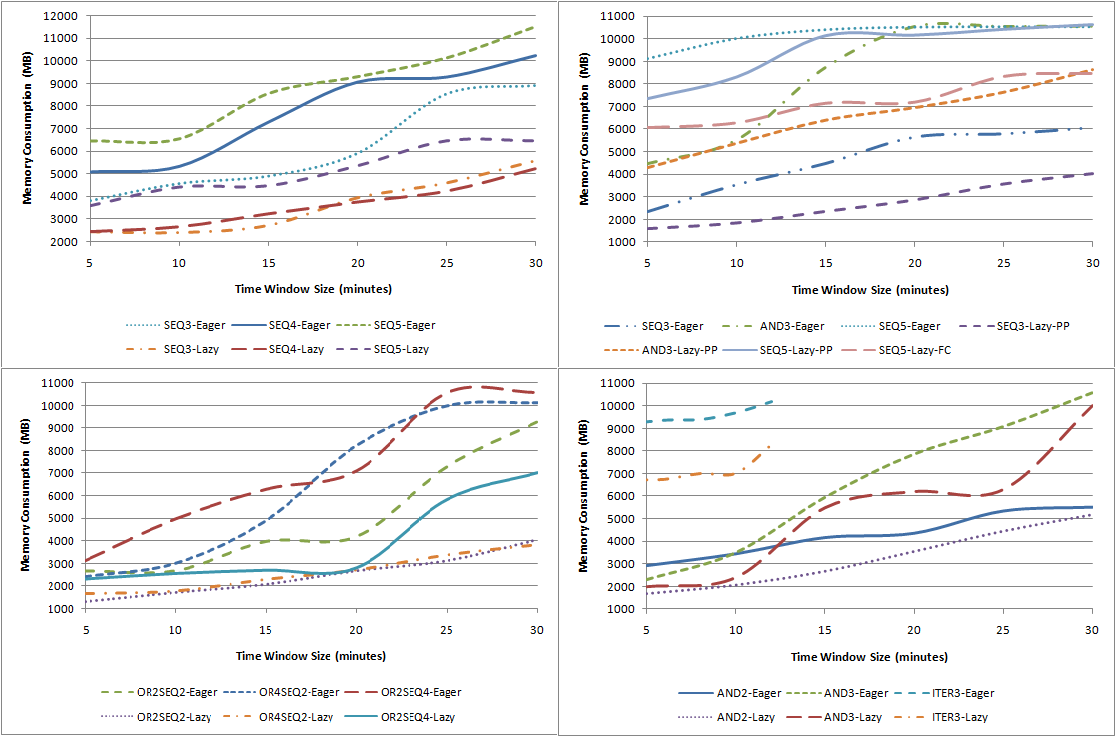}   \caption{Peak memory consumption as a function of time window size.} \label{fig:Peak-Memory} \end{figure*}

\begin{figure*}   \centering \includegraphics[width=.8\linewidth]{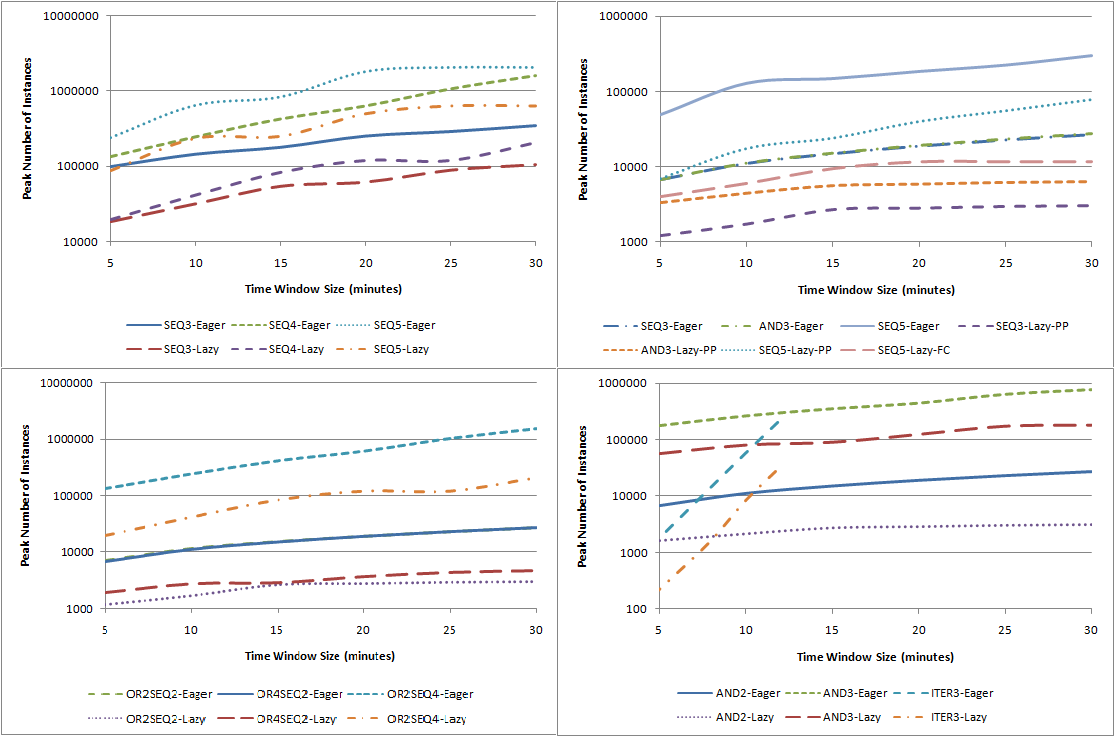}   \caption{Peak number of NFA instances as a function of time window size (logarithmic scale).} \label{fig:Peak-Instances} \end{figure*}

Memory consumption measurements are presented in Figures \ref{fig:Peak-Memory}
and \ref{fig:Peak-Instances}. Figure \ref{fig:Peak-Memory} compares
the total memory used during eager and lazy evaluation of various
patterns. For all pattern types, lazy automata require less memory
than eager automata. This is most evident for sequences and disjunctions,
for which the lazy solutions are 1.5 to 3 times more economical for
larger time windows. The smallest gain is observed for conjunctions,
which can be explained by the absence of ordering filters for this
pattern type (Section \ref{sub:Conjunction}).

It should be noted that the above figure only displays the peak memory
usage, i.e., the maximal recorded amount of required memory. However,
the difference in average memory consumption is more critical. Whereas
an eager NFA maintains the same high amount of reserved memory most
of the time, a Chain NFA only achieves its peak during brief 'bursts'.
These bursts follow an arrival of a rare event, which triggers a search
in the input buffer and creates a large number of intermediate instances.

Two main types of objects contribute to the overall memory consumption.
First, new NFA instances are created for each partial match. Second,
in lazy NFA primitive events are buffered upon arrival. While the
lazy evaluation method obviously keeps more buffered events, it creates
dramatically fewer instances. Figure \ref{fig:Peak-Instances} demonstrates
this by comparing the peak number of simultaneously active instances
during evaluation. Again, the logarithmic scale was used. On average,
a Lazy (Multi-)Chain NFA maintains 5 to 100 times fewer instances
at the same time. The biggest gain was achieved for large negation
and disjunction patterns.

\begin{figure*}   \centering \includegraphics[width=.8\linewidth]{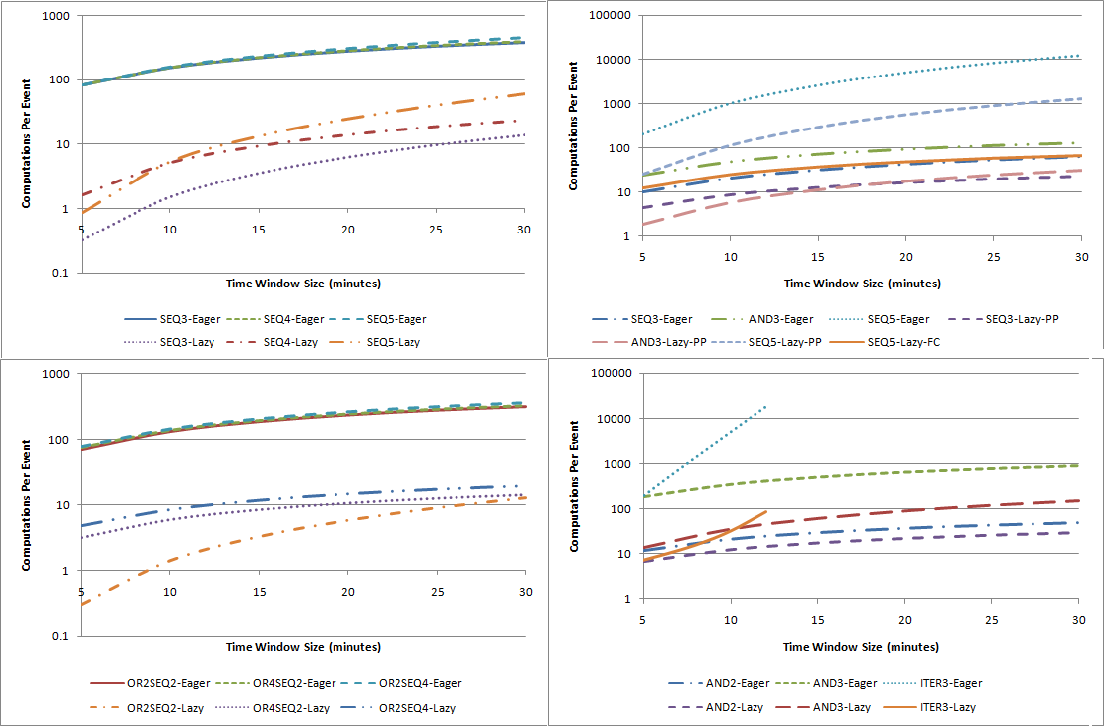}   \caption{Number of computations per event as a function of time window size (logarithmic scale).} \label{fig:Computations-Per-Event} \end{figure*}

\begin{figure*}   \centering \includegraphics[width=.8\linewidth]{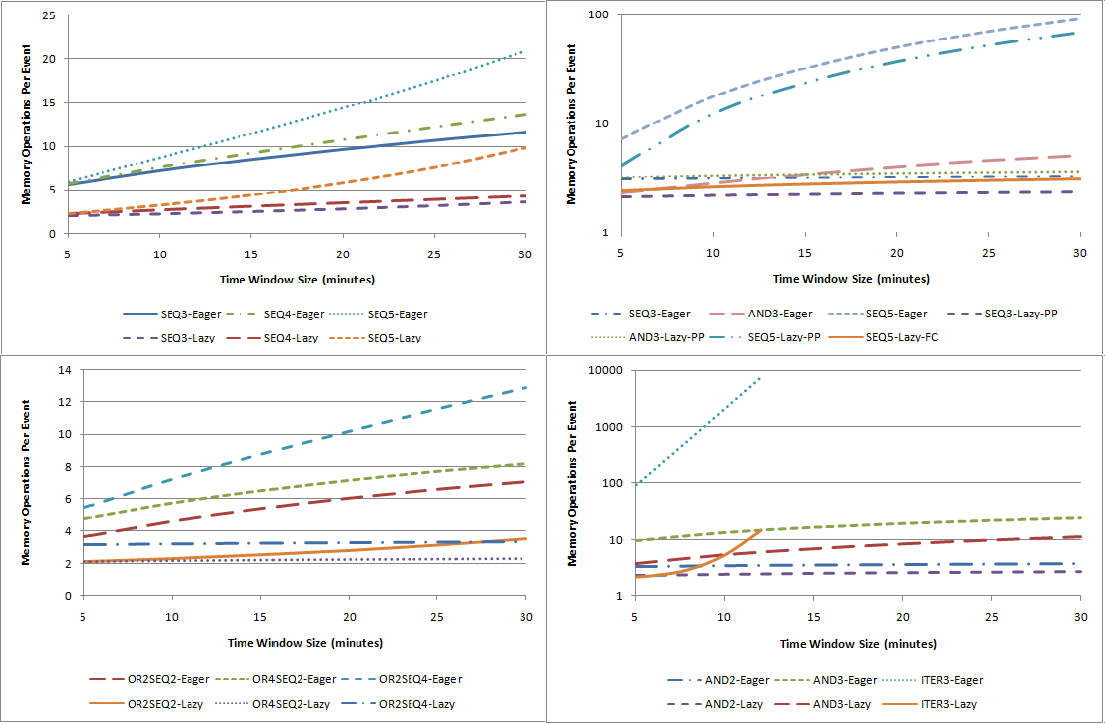}   \caption{Number of memory operations per event as a function of time window size (logarithmic scale for negations, conjunctions and iterations).} \label{fig:Memory-Operations-Per-Event} \end{figure*}

\begin{figure*}   \centering \includegraphics[width=.8\linewidth]{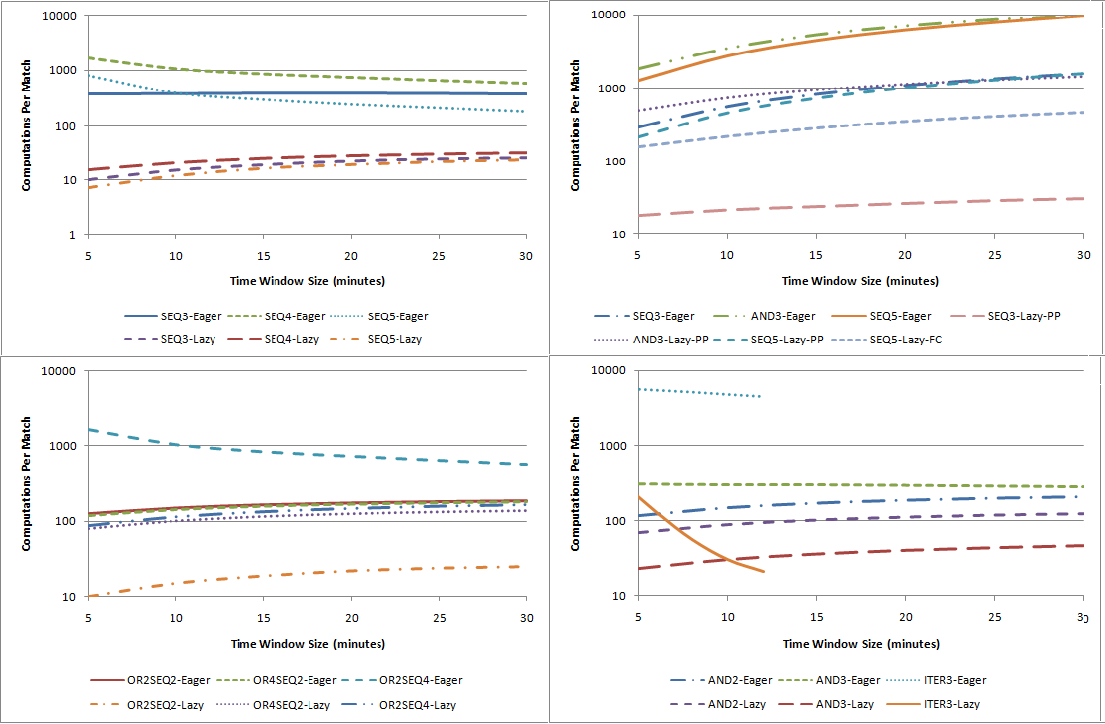}   \caption{Number of computations per pattern match as a function of time window size (logarithmic scale).} \label{fig:Computations-Per-Match} \end{figure*}

\begin{figure*}   \centering \includegraphics[width=.8\linewidth]{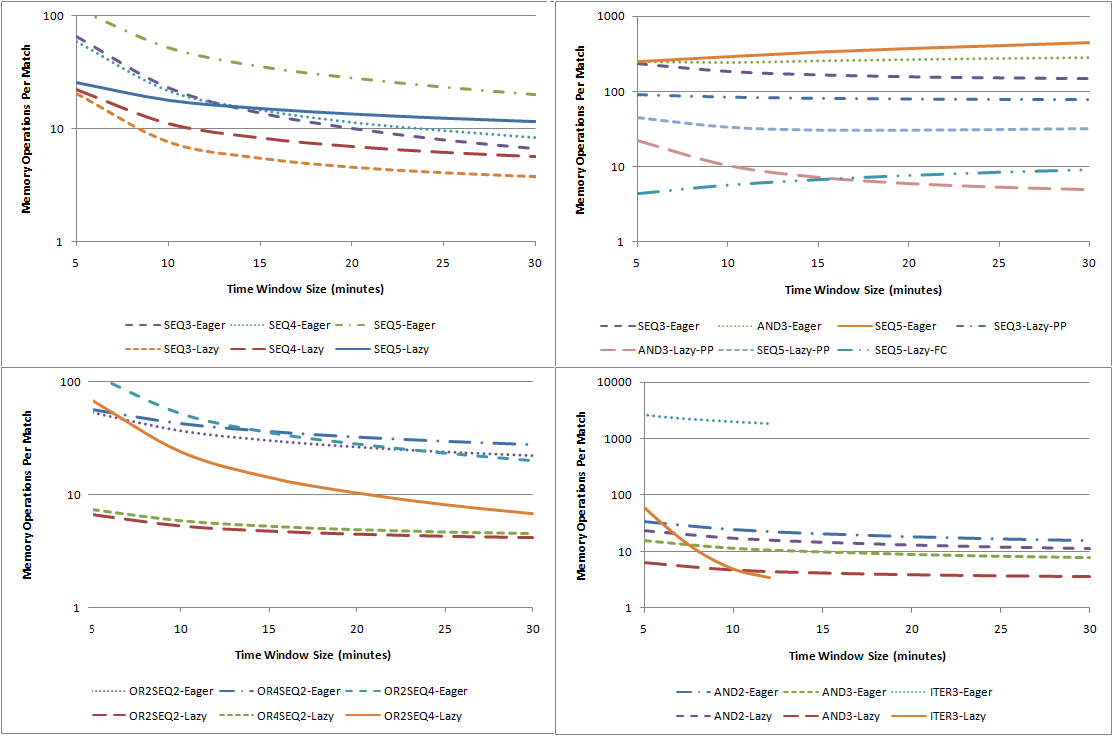}   \caption{Number of memory operations per pattern match as a function of time window size (logarithmic scale).} \label{fig:Memory-Operations-Per-Match} \end{figure*}

\begin{figure*}   \centering \includegraphics[width=.8\linewidth]{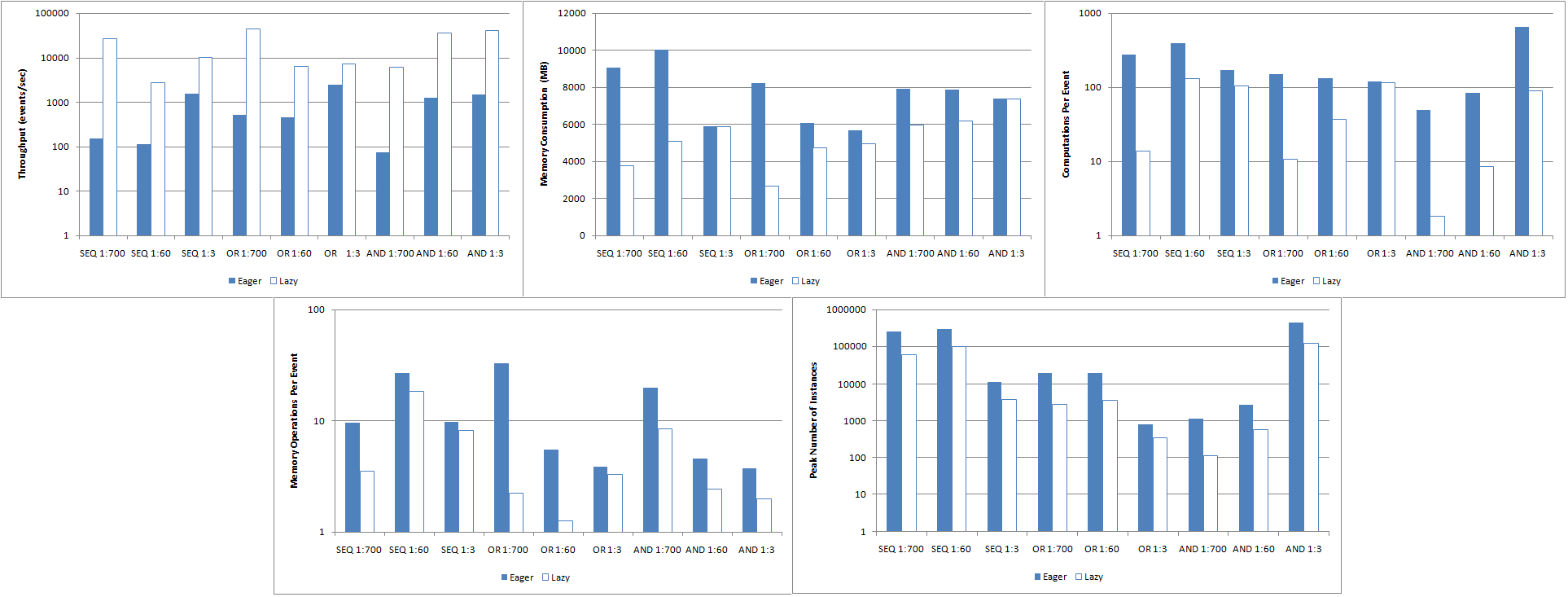}   \caption{Eager vs. Lazy evaluation performance for patterns with varied rarest-to-most-common event ratio: (a)throughput; (b)peak memory consumption; (c)number of predicate evaluations per event; (d)number of memory modifications per event; (e)peak number of simultaneously active instances.} \label{fig:Smallest-To-Largest-Event-Ratio} \end{figure*}

Now we proceed to the runtime complexity comparison. In this experiment,
we analyzed the two evaluation mechanisms in terms of operations performed
per primitive event and per successful pattern match. These operations
can be divided into two separate categories. The first includes the
calculations executed during predicate evaluation. Memory modifications,
such as NFA instance creation and deletion, buffering primitive events
and removing them upon evaluation, comprise the second category.

In Figure \ref{fig:Computations-Per-Event} the number of predicate
evaluations for each tested pattern can be seen. On average, lazy
NFAs execute 10 to 200 times fewer computations per primitive event.
This difference tends to be larger for more complex patterns involving
more event types.

Figure \ref{fig:Memory-Operations-Per-Event} presents the results
for memory modifications. Despite the need to maintain a complex data
structure for input buffering and to perform a large volume of insertions,
searches and removals, the lazy method does not suffer any performance
penalty in comparison to the eager approach. Furthermore, for most
patterns, significantly fewer operations were recorded during lazy
evaluation. This can be explained by the drastically reduced number
of instances to be created (and hence destroyed).

Finally, Figures \ref{fig:Computations-Per-Match} and \ref{fig:Memory-Operations-Per-Match}
display the runtime complexity measurements adjusted per detected
pattern match. The results are similar to those demonstrated above
for per event operations.

\subsection{Rarest-to-most-frequent Event Ratio}

\label{sub:Rarest-to-most-frequent-Event-Ra}

The core principle of the lazy evaluation mechanism is to re-order
the input stream so that less frequent events will be processed earlier.
Thus, from an analytical standpoint, this method achieves its biggest
performance gain when the frequencies of the participating events
are highly varied. As a consequence, the impact of this re-ordering
will diminish as events arrive at at more similar rates, and will
completely vanish if all event types are of identical frequency. An
interesting question is, how does a Chain NFA perform in these scenarios
as compared to an eager NFA?

In our next experiment we attempted to answer this question. To this
end, we evaluated a subset of patterns presented above on different
selections of event types, using a parameter called \textit{rarest-to-most-frequent
event ratio}. It is defined as the maximal ratio of arrival rates
of two event types in a pattern.

The frequency of each type was estimated according to the overall
number of companies belonging to the respective region. For example,
while NASDAQ has 267 registered companies located in Europe, there
are only 8 such companies in Africa. Hence, the ratio of the event
type \textit{AfrCompany} to \textit{EuCompany} is 8:267, or approximately
1:33. By carefully choosing sets of event types, we created a number
of patterns with a rarest-to-most-frequent event ratio ranging from
1:3 to 1:700. Each of the patterns was then evaluated with time window
size set to 20. All performance metrics presented in the beginning
of this section were measured during this experiment.

The results are displayed in Figure \ref{fig:Smallest-To-Largest-Event-Ratio}.
Note that in this case we are not interested in the absolute values.
Instead, we examine the performance difference of the eager and the
lazy evaluation mechanisms. As expected, this difference is inversely
proportional to the rarest-to-most-frequent ratio. For small values
of this ratio, the results for the lazy method are always at most
equal to those of the eager method.

\subsection{Chain NFA vs. Tree NFA on Dynamic Data}

In our next experiment we compared the performance of the NFAs discussed
above on data with dynamically changing frequencies of all primitive
events. For this experiment alone, synthetic data was used, generated
using the FINCoS framework \cite{FINCoS}. An artificial stream was
produced in which the rarest event was switched after each 100,000
incoming events. Then, all NFAs were tested against this input stream
using a simple sequence pattern of three event types. The number of
computations was measured after each 10,000 events.

\begin{figure}   \centering \includegraphics[width=\linewidth]{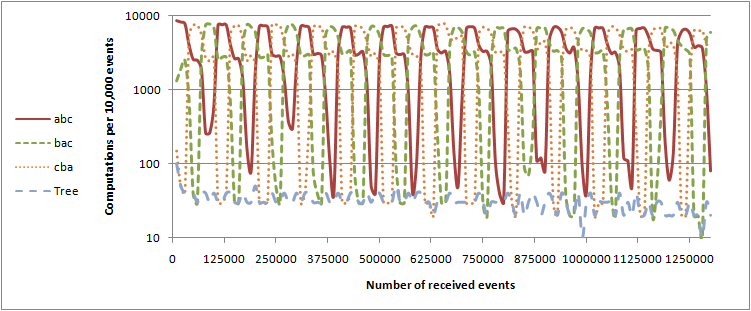}   \caption{Comparison of NFAs by number of operations on highly dynamic input (logarithmic scale) for sequence \textit{a,b,c}.} \label{fig:Dynamic-Performance} \end{figure}

Figure \ref{fig:Dynamic-Performance} demonstrates the results. Some
of the Chain NFAs were omitted due to very similar results. The x-axis
represents the number of events from the beginning of the stream.
It can be thought of as the closest estimate to the time axis. The
y-axis represents the number of computations per 10,000 events.

This figure illustrates the superiority of the Tree NFA over its competitors
and its high adaptivity to changes in event arrival rates. At any
single point there is one frequency order that is the most efficient
given the current event frequencies. The performance gain of the Chain
NFA based on that order over the other Chain NFAs reaches up to two
orders of magnitude. However, as soon as the event frequencies change,
this NFA loses its advantage. On the other hand, the Tree NFA shows
consistent improvement over all Chain NFAs regardless of the input
arrival rates.

\subsection{State-of-the-art Comparison}

\label{sub:State-of-the-art-Comparison}

In our last experiment we compared the performance of the lazy Chain
NFA with the official implementation of SASE+\cite{ZhangDI10}. The
set-up described in Section \ref{sub:Time-Window-Size} was used.
Sequence, negation and iteration patterns were employed. To avoid
modifying the code of SASE+, we only measured throughput during the
evaluation.

\begin{figure}   \centering \includegraphics[width=.5\linewidth]{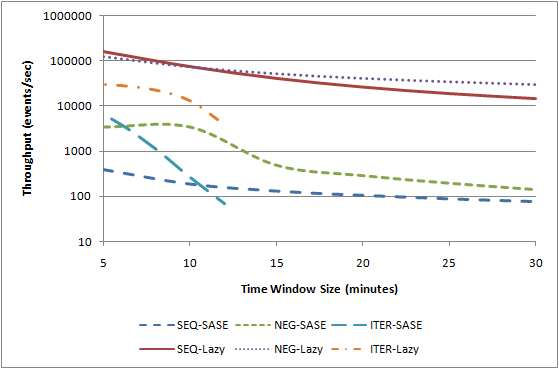}   \caption{Throughput of SASE+ and the lazy Chain NFA for sequence, negation and iteration patterns (logarithmic scale).} \label{fig:SASE-Comparison} \end{figure}

Figure \ref{fig:SASE-Comparison} shows the results. Since our implementation
of the eager NFA was based on SASE+ and carefully followed its formal
description, a high degree of similarity with Figure \ref{fig:Throughput}
can be observed. As demonstrated earlier, using the Lazy Evaluation
method yields a performance gain of one to two orders of magnitude
in terms of processed events per time unit.

\section{Related Work}

\label{sec:Related-Work}

The detection of complex events over streams has become a very active
research field in recent years \cite{CugolaM12}. The earliest systems
designed for solving this problem fall under the category of Data
Stream Management Systems. Those systems are based on SQL-like specification
languages and focus on processing data coming from continuous, usually
multiple input streams. Examples include NiagaraCQ \cite{ChenDTW00},
TelegraphCQ \cite{ChandrasekaranDFHHKMRRS03}, Aurora \cite{BalakrishnanBCCCCGSSTTZ04}
and STREAM \cite{ilprints583}. Later, the need to analyze event notifications
of interesting situations -- as opposed to generic data -- was identified.
Then, \textit{complex event processing systems} were introduced. One
example of an advanced CEP system is Amit \cite{AdiE04}, based on
a strongly expressive detection language and capable of processing
notifications received from different sources in order to detect patterns
of interest. SPADE \cite{GedikAWYD08} is a declarative stream processing
engine of System S. System S is a large-scale, distributed data stream
processing middleware developed by IBM. It provides a computing infrastructure
for applications that need to handle large scale data streams. Cayuga
\cite{BrennaDGHOPRTW07,DemersGHRW06,DemersJB07} is a general purpose,
high performance, single server CEP system developed at Cornell University.
Its implementation focuses on multi-query optimization methods.

Apart from SASE \cite{WuDR06} and SASE+ \cite{AgrawalDGI2008}, thoroughly
discussed above, many other event specification languages were proposed.
CEL \cite{DemersGHRW06,DemersJB07} is a declarative language used
by the Cauyga system. It supports patterns with Kleene closure and
event selection strategies. CQL \cite{ArasuBW06} made it possible
to create transformation rules with a unified syntax for processing
both information flows and stored relations. TESLA \cite{CugolaM10}
combines high expressiveness with a relatively small set of operators.
It offers completely programmable per-event selection policies. CEDR
\cite{BargaGAH07} presents a complex and flexible temporal model,
based on three different timings. Our lazy evaluation framework is
based on SASE+ \cite{GyllstromADI08}, an expressive language, including
constructs such as iterations and aggregations. Nevertheless, the
operators it supports serve as basic blocks for most of the described
languages.

Optimizing event processing performance by pattern re-writing (re-ordering)
techniques has been discussed in numerous studies. \cite{RabinovichEG11}
and \cite{Schultz-MollerMP09} employ formally defined cost models
to derive the best possible evaluation plan, according to which input
patterns are re-organized. The objectives of query re-writing are
throughput and latency balancing. Both frameworks consider distributed
CEP environment.

The concept of lazy evaluation has been widely applied in multiple
research fields. Related work was also conducted in the context of
complex event processing. ZStream \cite{MeiM09} uses tree-based query
plans to represent complex patterns. As the most efficient plan is
chosen, the pattern detection order is dynamically adjusted by internal
buffering. In \cite{AkdereMCT08}, the authors describe ``plan-based
evaluation'', where temporal properties of primitive events are exploited
to reduce network communication costs. \cite{ChanFGR02} presents
an XPath-based mechanism for filtering XML documents in stream environments.
This method postpones costly operations as long as possible. While
many of the ideas discussed by the aforementioned studies are close
to ours, none considers the finite automata environment, which is
the primary focus of our work.

Several authors propose postponing and preprocessing mechanisms for
NFA-based evaluation frameworks. In \cite{ZhangDI2014}, the authors
present an optimization method based on sharing common operations
between instances and postponing per-instance part. The focus of that
technique is on improving the performance of Kleene closure patterns
under the skip-till-any-match selection policy. \cite{DoussonM07}
discusses a mechanism similar to ours, including the concept of storing
incoming events and evaluating more selective events before more frequent
ones. However, the authors do not consider complex pattern types,
such as negation and iteration. \cite{CadonnaGB12} proposes a matching
strategy including a preprocessing step, which is similar to our input
buffering. It is used for applying pruning techniques, such as filtering
and partitioning, rather than for event reordering. APAM \cite{YiLW16}
is a hybrid eager-lazy evaluation method. For each event type, it
determines which method to use for minimizing detection latency.

The concept of lazy evaluation has also been proposed in the related
research field of online processing of XML streams. \cite{ChanFGR02}
describes an XPath-based mechanism for filtering XML documents in
stream environments. This mechanism postpones costly operations as
long as possible. However, the goal in this setting is only to detect
the presence or absence of a match, whereas our focus is on finding
all possible matches between primitive events. In \cite{GreenGMOS04},
a technique for lazy construction of a DFA (Deterministic Finite Automaton)
on-the-fly is discussed. This work is motivated by the problem of
exponential growth of automata for XPath pattern matching. Our work
solves a different problem of minimizing the number of runtime NFA
instances rather that the size of the automaton itself. In addition,
while there is some overlap in the semantics of CEP and XPath queries,
they were designed for different purposes and allow different types
of patterns to be defined.

The lazy evaluation mechanism for NFA-based complex event processing
was originally proposed in \cite{KolchinskySS15}. It introduced the
main ideas and demonstrated how this method can be applied on simple
sequence patterns. However, other operator types and composite patterns
were not addressed, nor was the correctness of the presented constructions
formally proved. On the contrary, this paper is the first to provide
a universal solution, suitable for evaluating complex patterns containing
conjunction, disjunction, negation, and iteration operators, as well
as the corresponding proof of correctness.

\section{Conclusions and Future Work}

\label{sec:Conclusions}

This paper presented a lazy evaluation mechanism for efficient detection
of complex event patterns. Unlike previous solutions, our system does
not process the events in order of their arrival, but rather according
to their ascending order of frequency. Two NFA topologies were proposed
to implement the above concept. The Chain NFA requires the arrival
rates of the events in the pattern to be known in advance. The Tree
NFA utilizes an adaptive approach by computing the actual frequency
order on-the-fly. Our experimental results showed that both Chain
NFA and Tree NFA achieve significant improvement over the eager evaluation
mechanism in terms of throughput, memory consumption, and runtime
complexity. Our future work will explore several research avenues,
such as efficient multi-query support, handling uncertainty, and adapting
our techniques to a distributed environment.

\bibliographystyle{plain}
\bibliography{references}

\end{document}